\documentclass[hyper,a4paper]{JHEP3}
\usepackage{epsfig}
\usepackage{latexsym}
\usepackage{graphicx}
\usepackage{amsmath}
\usepackage{amsfonts}   
\usepackage{amssymb}    
\usepackage{float}
\usepackage{bm}
\newcommand{\comment}[1]{}
\newcommand{\newc}{\newcommand}

\def\thetab0{\theta_{B_0}}

\def\r2{\sqrt 2}
\def\beq{\begin{equation}}
\def\eeq{\end{equation}}
\def\bea{\begin{eqnarray}}
\def\eea{\end{eqnarray}}
\def\sinW2{\sin^2\theta_W}

\def\mz2{M_{z}^2}
\def\c2b{\cos 2\beta}

\def\mz{M_Z}

\def\sec2w{sec^2\theta_W}

\def\gmin2{(g-2)_\mu}

\catcode`\@=11 

\def\lsim{\mathrel{\mathpalette\@versim<}}
\def\gsim{\mathrel{\mathpalette\@versim>}}
\def\@versim#1#2{\vcenter{\offinterlineskip
    \ialign{$\m@th#1\hfil##\hfil$\crcr#2\crcr\sim\crcr } }}

\newc{\ra}{\rightarrow}
\newc{\s}{\smallskip}
\newc{\nn}{\noindent}
\newc{\non}{\nonumber}

\def \chonep{{\wt\chi_1}^{+}}
\def \chonem{{\wt\chi_1^-}}
\def \chonep2{{\wt\chi_2^+}}
\def \chonem2{{\wt\chi_2^-}}

\title
{Neutrino masses and mixing, lightest neutralino decays and a solution 
to the $\mu$ problem in supersymmetry}
\author{Pradipta Ghosh and Sourov Roy\\
Department of Theoretical Physics and Centre for Theoretical
Sciences,\\
 Indian Association for the Cultivation of Science, \\
2A $\&$ 2B Raja S.C. Mullick Road, Kolkata 700 032, India\\
E-mails:\email{ tppg@iacs.res.in,  tpsr@iacs.res.in
}\\}

\abstract{We examine in detail the neutrino masses and mixing patterns in 
an extension of the minimal supersymmetric standard model with three 
gauge-singlet neutrinos and R-parity violation. The Majorana masses for the 
gauge-singlet neutrinos as well as the usual $\mu$-term for the Higgs 
superfields are generated at the electroweak scale through the vacuum 
expectation values of the singlet sneutrinos. The resulting effective mass 
matrix for the three light neutrinos have contributions from the seesaw 
mechanism involving the singlet neutrinos as well as due to the mixing with 
the heavy neutralinos. This model is popularly known in the literature as the
``$\mu$ from $\nu$ supersymmetric standard model" ($\mu\nu$SSM). We show that 
even with flavour diagonal neutrino Yukawa couplings, the global data on 
three-flavour neutrinos can be well accounted for in this scenario, at the 
tree level. We also analyze the mixing in the chargino and the Higgs sector 
and calculate the decays of the lightest supersymmetric particle in this 
model. The decay branching ratios show certain correlations with the neutrino 
mixing angles, which can be tested at the LHC. Some other phenomenological 
implications of such a model have been discussed.}

\keywords{Supersymmetry Phenomenology}

\begin{document}
\section{Introduction}
\label{Introduction}
The experimental results on neutrinos provide strong evidence in favour of 
non-zero neutrino masses and mixing angles. Various neutrino oscillation
experiments suggest that the mixing pattern of the three light neutrinos is
bilarge, that is to say there are two large mixing angles and one small 
mixing angle. The data can be explained very well with the following set of 
mass squared differences and mixing angles 
\cite{boris_review_08, Schwetz-Valle}
\bea
7.05 \times 10^{-5} {\rm eV}^2 \leq \Delta m^2_{21} \leq 8.34 \times 10^{-5} 
{\rm eV}^2 ~~~~~~~~(3\sigma),
\label{eq-msq21}
\eea
\bea
2.07 \times 10^{-3} {\rm eV}^2 \leq |\Delta m^2_{31}| \leq 2.75 \times 
10^{-3} {\rm eV}^2 ~~~~~~(3\sigma),
\label{eq-msq31}
\eea
\bea
30^\circ \leq {\theta_{12}} \leq 37^\circ,~ 
35^\circ \leq {\theta_{23}} \leq 54^\circ,~
{\theta_{13}} \leq 13^\circ ~~~~~~(3\sigma), 
\label{eq-mixing-angles}
\eea
where $\Delta m^2_{ij} \equiv m^2_i - m^2_j$.  

In order to explain the results in eqs.(\ref{eq-msq21})--
(\ref{eq-mixing-angles}), one needs to go to a theory beyond the standard 
model (SM). A very interesting possibility to look for new physics is 
supersymmetry (SUSY) which has the ability to provide a solution of the 
so-called `gauge hierarchy problem' connected with the mass of the Higgs 
boson. SUSY predicts new particles at the TeV scale which can be tested at
the forthcoming Large Hadron Collider (LHC). Naturally it is tempting to see
whether TeV scale SUSY is also one of the candidates which can explain the 
observed pattern of neutrino mass squared differences and mixing. There have 
been several proposals in recent times, which attempt to explain the 
experimental data on neutrinos in the context of a supersymmetric theory.
Perhaps the most well studied class of models in this context is the one 
which includes R-parity violation 
\cite{r-parity-mssm, r-parity-mssm-neutrino1} in the Minimal Supersymmetric 
Standard Model (MSSM). Neutrino mass squared differences and mixing angles 
have been calculated under various assumptions and it has been found that 
the neutrino data can be explained well when contributions to the neutrino 
mass matrix at the tree and one-loop level are considered 
\cite{r-parity-mssm-neutrino2,r-parity-review}. 

On the other hand, though SUSY provides some elegant solutions to accommodate
the experimental data on neutrinos, it has been plagued by a few urgent 
questions which do not have very satisfactory answers yet. One of them is the
so-called ``$\mu$-problem"\cite{Kim-Niles} related to the bilinear term 
$\mu {\hat H}_1 {\hat H}_2$ in the MSSM superpotential. The electroweak 
symmetry breaking requires the value of $\mu$ to be roughly of the order 
of a few hundreds of GeV. This requires, in the absence of any fine 
cancellation, that $\mu$ is roughly of the order of the soft scalar masses 
and both of them should be somewhere around a TeV or a few hundreds of GeV. 
Since $\mu$ parameter respects SUSY, there is no obvious reason why it should 
be of the same order as SUSY breaking soft scalar masses. This defines 
the ``$\mu$-problem". There have been several attempts to address the solution 
to this problem and all of them requires the vacuum expectation value(s) 
(VEVs) of some additional field(s) to generate the $\mu$-term after the 
symmetry breaking. 

One of the solutions is the next-to-minimal supersymmetric standard model 
(NMSSM), which introduces a superfield $\hat S$, singlet under the SM gauge 
group. The $\mu$ term is absent from the superpotential and it arises when 
the scalar component of $\hat S$ acquires a VEV. This VEV is determined in 
terms of the soft supersymmetry breaking terms through the minimization 
condition. If the SUSY breaking scale is close to the electroweak (EW) scale 
then the effective $\mu$ term is also of the order of the EW scale. However,
as in the case of MSSM, the NMSSM also cannot explain the observed pattern of
neutrino masses and mixing. 

It is also important to note in this context, that another very well 
known mechanism of generating small neutrino masses and bilarge mixing angles 
in a SUSY model, compatible with the experimental data, is the seesaw 
mechanism which introduces gauge singlet neutrino superfields. In such cases 
the MSSM superpotential contains additional terms involving the Yukawa 
couplings for neutrinos as well as Majorana masses for the gauge 
singlet neutrinos. In the conventional scenario, the neutrino Yukawa
couplings are assumed to be $\sim {\cal O}$(1), whereas the Majorana masses for 
the gauge singlet neutrinos are taken somewhere around $10^{14}$ GeV or so. 
In this case light neutrino masses as small as $10^{-2}$ eV can be generated.
One viable alternative to the usual seesaw mechanism is to consider the
{\em TeV-scale} seesaw. This possibility is very interesting since it may 
provide a direct way to probe the gauge singlet neutrino mass scale at the
LHC and does not need to introduce a very high energy scale in the theory. 
However, in order to generate small active neutrino masses one needs to 
consider neutrino Yukawa couplings to be of order $10^{-6}$. This choice is 
reasonable since we know that the electron Yukawa coupling should also be of 
the order of $10^{-6}$. 

In this work we study in details, a model of neutrino masses and mixing which 
introduces the gauge singlet neutrino superfields (${\hat\nu}^c_i$) to solve 
the $\mu$ problem in a way similar to that of NMSSM. The terms in the 
superpotential involving the ${\hat\nu}^c_i$ include the neutrino 
Yukawa couplings, the trilinear interaction terms among the singlet neutrino 
superfields as well as a term which couples the Higgs superfields to the 
$\hat\nu^c_i$. In addition, there are corresponding soft SUSY breaking terms in 
the scalar potential. When the scalar components of ${\hat\nu}^c_i$ get VEVs 
through the minimization conditions of the scalar potential, an effective 
$\mu$ term with an EW scale magnitude is generated \cite{munoz-lopez_fogliani}.
In addition, small Majorana masses of the active neutrinos are generated due 
to the mixing with the neutralinos as well as due to the seesaw mechanism 
involving the gauge singlet neutrinos. In such a scenario, we aim to explain 
the experimental data on neutrinos discussed in the beginning. In particular, 
we show that it is possible to provide a theory of neutrino masses and mixing 
explaining the experimental data, even with a flavour diagonal neutrino Yukawa 
coupling matrix, without resort to an arbitrary flavour structure in the 
neutrino sector. This essentially happens because of the mixing involved in 
the neutralino-neutrino (both doublet and singlet) system mentioned above. 
We perform a detailed analytical and numerical work and show 
that the three flavour neutrino data can be accommodated in such a scenario. 
In addition, we observe that in this model different neutrino mass hierarchies 
can also be obtained with correct mixing pattern, at the tree level. 

In this model the neutral Higgs bosons mix with the sneutrinos and charged 
Higgs bosons mix with the charged sleptons. The corresponding scalar, 
pseudoscalar and charged scalar mass squared matrices are enlarged and we take 
into account the constraints on the parameters coming from the positivity of 
the squared scalar masses. In the fermionic sector, in addition to 
the neutralino-neutrino mixing there is mixing also between the charginos and 
the charged leptons. This can also have implications for phenomenological 
studies, particularly in the context of future colliders. Because of the mixing 
between the neutralinos and the neutrinos, the lightest neutralino, which is 
the lightest superparticle (LSP) for most of the parameter space, can have 
novel decay modes which can be correlated with the neutrino mixing pattern. 
This can provide some unique signatures of such a scenario which can be tested 
at the LHC.

As we have mentioned earlier, that in order to get the correct light neutrino 
mass scales, the neutrino Yukawa couplings should be of the order of $10^{-6}$ 
or so. This is because the TeV scale VEVs of the singlet neutrinos induce
Majorana mass terms of themselves also at the TeV scale. Similarly, the
neutralino-neutrino mixing provides correct light neutrino mass scales only
when the Yukawa couplings of the neutrinos are somewhere close to that of the
electron. This model has been named as the ``$\mu \nu$SSM" in 
ref.\cite{munoz-lopez_fogliani}. Thus in this model, an interesting proposal
has been given, where the generation of small neutrino masses and the solution 
to the $\mu$ problem can be accommodated with the same set of gauge singlet 
neutrino superfields without introducing an extra singlet such as in the case 
of NMSSM. The spectrum and parameter space of 
this model, with three gauge singlet neutrino superfields, were discussed in
\cite{munoz-lopez-2}. However, a detailed discussion of the issue of neutrino masses and bilarge neutrino mixing, in order to accommodate the three flavour neutrino data, has been lacking and that is what we want to provide in the present work. We would like to emphasize here that our analysis shows that even with flavour diagonal neutrino Yukawa couplings, the resulting structure of the 
effective mass matrix of the light neutrinos can explain the bilarge pattern 
of mixing. In addition, we explore the correlation between neutrino mixing 
and the decay pattern of the LSP in this model and discuss some other 
interesting phenomenological implications.

Other models which address the neutrino experimental data and the $\mu$ 
problem are essentially extensions of NMSSM. One of these proposals
\cite{kitano-oda} include both the gauge singlet neutrino superfields 
(${\hat \nu}^c_i$) and the extra singlet ($\hat S$) of the NMSSM. In this 
case the Majorana masses of the singlet neutrinos are also generated at the 
EW scale through the VEV of the scalar component of $\hat S$. R-parity may 
be broken spontaneously and the light neutrino masses are generated through 
the seesaw at the EW scale. Another possibility is discussed in 
ref.\cite{chemtob-pandita}, where the effective bilinear R-parity violating 
terms are generated through the VEV of the singlet scalar $S$. Naturally, in 
this case only one neutrino mass is generated at the tree level whereas the 
other two masses are generated at the loop level. In 
\cite{gautam-moreau-abada}, two neutrino masses
are generated at the tree level by including explicit bilinear R-parity
violating terms in addition to the R-parity breaking term involving $\hat S$. 

The plan of the paper is the following. We start with the description of the
model in Sec. II by writing down the superpotential and the soft supersymmetry
breaking interaction terms. We also derive the minimization equations of the 
scalar potential and discuss some issues related to the vacuum expectation
values of the left sneutrinos. In Sec. III we continue our discussion of the
scalar sector with a more detailed look. Sec. IV describes the fermionic 
sector of the model where neutralinos mix with both the doublet and singlet 
neutrinos and the charginos mix with the charged leptons. In Sec. V we provide 
a detailed discussion of the effective mass matrix of the three light 
neutrinos, arising because of the neutralino-neutrino mixing. Analytical 
expressions of the mass eigenvalues and eigenvectors are derived under certain 
conditions, using degenerate perturbation theory. We construct the neutrino 
mixing matrix and show that it is possible to have two large and one small 
mixing angles in general. A detailed numerical analysis has been performed and 
we compare our results with that obtained using the approximate analytical 
formulae. We show that for realistic parameter choices, it is possible to fit 
the three flavour global neutrino data in this scenario, even with a flavour 
diagonal neutrino Yukawa coupling matrix. The decays of the lightest 
supersymmetric particle are discussed in Sec. VI and it has been observed that 
certain decay branching ratios are correlated with the neutrino mixing angles. 
We make concluding remarks in Sec. VII with possible future directions of our 
work. The details of various scalar mass squared matrices and the Feynman 
rules for the LSP decay calculations are relegated to the appendices.


\section{The model and its minima}
\label{The model and its minima}
\subsection{Superpotential}
In this section we introduce the model along the lines of 
ref.\cite{munoz-lopez_fogliani}, discuss its basic features and set our
notations. We introduce three gauge singlet neutrino superfields, 
${\hat \nu}_i^c$ ($i = e, {\mu}, {\tau}$), in addition to the fields of the 
minimal supersymmetric standard model. The superpotential of the model is 
written as  
\bea
W &=& \epsilon_{ab}(Y^{ij}_u\hat H^b_2\hat Q^a_i\hat u^c_j + Y^{ij}_d\hat H^a_1
\hat Q^b_i\hat d^c_j  + Y^{ij}_e\hat H^a_1\hat L^b_i\hat e^c_j + Y^{ij}_\nu 
\hat H^b_2\hat L^a_i\hat \nu^c_j)\nonumber \\
&-&\epsilon_{ab} \lambda^i\hat \nu^c_i\hat H^a_1\hat H^b_2
+ \frac{1}{3}\kappa^{ijk}\hat \nu^c_i\hat \nu^c_j\hat \nu^c_k,
\label{superpotential}
\eea
where $\hat H_1$ and $\hat H_2$ are the down-type and up-type Higgs 
superfields, respectively. The $\hat Q_i$ are doublet quark superfields, 
${\hat u}^c_j$ [${\hat d}^c_j$] are singlet up-type [down-type] quark 
superfields. The $\hat L_i$ are the doublet lepton superfields, and the 
${\hat e}^c_j$ are the singlet charged lepton superfields. Here $a,b$ are SU(2) 
indices, and $\epsilon_{12}$ = --$\epsilon_{21}$ = 1. Note that the usual 
bilinear $\mu$-term of the MSSM is absent from the superpotential whereas 
two additional trilinear terms are introduced involving the Higgs superfields, 
$\hat H_1$ and $\hat H_2$, and the three gauge singlet neutrino superfields, 
$\hat \nu^c_i$. This is done by imposing a $Z_3$ symmetry which is also used 
in the case of NMSSM. If the scalar potential of the model is such that 
non-zero vacuum expectation values of the scalar components ($\tilde \nu^c_i$) 
of the singlet superfields $\hat \nu^c_i$ are induced, an effective bilinear 
term $\mu\hat H^a_1\hat H^b_2$ is generated, where the coefficient $\mu 
\equiv \lambda^i \langle \tilde \nu^c_i \rangle$. In the presence of soft 
supersymmetry breaking, it is usually expected that the VEVs of 
$\tilde \nu^c_i$ are at the electroweak scale. Hence the value of $\mu$ is 
of the order of the electroweak scale as long as the dimensionless 
couplings $\lambda_i$ are $\sim \cal O$(1). This gives us a solution to the 
so-called ``$\mu$-problem". The last term in the superpotential with the 
coefficient $\kappa^{ijk}$ is included in order to avoid an unacceptable axion 
associated to the breaking of a global $\rm U(1)$ symmetry 
\cite{ellis-gunion-haber}. This term generates effective Majorana masses for 
the singlet neutrinos at the electroweak scale. 

The last two terms in (\ref{superpotential}) explicitly break lepton number 
(L) and hence the R-parity, which is defined by $R = (-1)^{L + 3B + 2s}$. Here 
B is the baryon number and s is the spin. Note that $R = +1$ for particles 
and $-1$ for superpartners. Since lepton number is explicitely broken, 
no unwanted massless Majoron appears in this model. One should also notice 
that the term in the superpotential involving the neutrino Yukawa couplings 
$Y^{ij}_\nu$, generate effective bilinear R-parity violating interactions 
$\epsilon^{i} \hat H_2\hat L_i$. Here $\epsilon^{i}$ is determined in terms 
of the VEVs of ${\tilde \nu}^c_i$ and is given by $\epsilon^{i} = Y_{\nu}^{ij} 
\langle \tilde \nu^c_{j} \rangle$. R-parity breaking implies that the lightest
supersymmetric particle (LSP) is not stable and it cannot be a candidate for 
dark matter. The decay of the LSP may produce some interesting signatures at 
the LHC, which can have certain correlations with the neutrino oscillation 
data. In addition, one can measure displaced vertices originating from the LSP
decay.  

It should be mentioned here that neutrino masses and bilarge neutrino mixing 
have also been studied in an R-parity violating supersymmetric theory with 
gauge singlet neutrino superfields \cite{biswarup-srikanth}. However, in that 
analysis terms of the type ${\hat \nu}^c {\hat H}_1 {\hat H}_2$ have been 
dropped from the superpotential because of very small coefficient. Analysis 
has also been carried out in the context of the observed baryon asymmetry of 
the Universe \cite{farzan-valle}. Finally, one should note that the discrete 
$Z_3$ symmetry of the superpotential is spontaneously broken in the vacuum. 
This might lead to cosmological domain wall problem \cite{domain-wall}. 
However, the solutions to this problem exist \cite{domain-wall-soln} and will 
also work in this case.


\subsection{Soft terms}
Let us now specify the soft-supersymmetry-breaking terms of this model. 
We will confine ourselves in the framework of supergravity mediated 
supersymmetry breaking. The Lagrangian $\mathcal{L}_{\text{soft}}$, 
containing the soft-supersymmetry-breaking terms is given by

\bea
-\mathcal{L}_{\text{soft}} &=&
(m_{\tilde{Q}}^2)^{ij} {\tilde Q^{a^*}_i} \tilde{Q^a_j}
+(m_{\tilde u^c}^{2})^{ij}
{\tilde u^{c^*}_i} \tilde u^c_j
+(m_{\tilde d^c}^2)^{ij}{\tilde d^{c^*}_i}\tilde d^c_j
+(m_{\tilde{L}}^2)^{ij} {\tilde L^{a^*}_i}\tilde{L^a_j} \nonumber \\
&+&(m_{\tilde e^c}^2)^{ij}{\tilde e^{c^*}_i}\tilde e^c_j 
+ m_{H_1}^2 {H^{a^*}_1} H^a_1 + m_{H_2}^2 {H^{a^*}_2} H^a_2 +
(m_{\tilde{\nu}^c}^2)^{ij}  {\tilde{\nu}^{c^*}_i} \tilde\nu^c_j \nonumber \\
&+& \epsilon_{ab} \left[
(A_uY_u)^{ij} H_2^b\tilde Q^a_i \tilde u_j^c +
(A_dY_d)^{ij} H_1^a \tilde Q^b_i \tilde d_j^c +
(A_eY_e)^{ij} H_1^a \tilde L^b_i \tilde e_j^c + \text{H.c.}  \right] 
\nonumber \\
&+&\left[\epsilon_{ab}(A_{\nu}Y_{\nu})^{ij} H_2^b \tilde L^a_i \tilde 
\nu^c_j-\epsilon_{ab} (A_{\lambda}\lambda)^{i} \tilde \nu^c_i H_1^a  H_2^b+
\frac{1}{3} (A_{\kappa}\kappa)^{ijk} \tilde \nu^c_i \tilde \nu^c_j \tilde 
\nu^c_k\ + \text{H.c.} \right] \nonumber \\
&-& \frac{1}{2} \left(M_3 \tilde{\lambda}_3 \tilde{\lambda}_3
+ M_2 \tilde{\lambda}_2 \tilde{\lambda}_2 + M_1 \tilde{\lambda}_1 
\tilde{\lambda}_1 + \text{H.c.} \right).
\label{Lsoft}
\eea

In eq.(\ref{Lsoft}), the first two lines consist of squared-mass terms
of squarks, sleptons (including the gauge singlet sneutrinos ${\tilde \nu}^c_i$)
and Higgses. The next two lines contain the trilinear scalar couplings. 
Finally, in the last line of eq.(\ref{Lsoft}), $M_3, M_2$, and $M_1$ are 
Majorana masses corresponding to $SU(3)$, $SU(2)$ and $U(1)$  gauginos 
$\tilde{\lambda}_3, \tilde{\lambda}_2$, and $\tilde{\lambda}_1$,  respectively. 
The tree-level scalar potential receives the usual D and F term contributions,
in addition to the terms from $\mathcal{L}_{\text{soft}}$. 


\subsection{The neutral scalar potential and the electroweak symmetry breaking 
conditions}

We assume that only the neutral scalar fields develop in general the following 
vacuum expectation values while minimizing the scalar potential$\colon$
\bea
\langle H_1^0 \rangle = v_1 \, , \quad
\langle H_2^0 \rangle = v_2 \, , \quad
\langle \tilde \nu_i \rangle = \nu_i \, , \quad
\langle \tilde \nu_i^c \rangle = \nu_i^c.
\label{vevs}
\eea
The tree level neutral scalar potential looks like 

\bea
\langle V_{\text{neutral}}\rangle &=& |\sum_{i,j}Y^{ij}_{\nu}{\nu_i}{\nu^c_j}-
\sum_{i}\lambda^i{\nu^c_i} v_1|^2\ + \sum_{j}|\sum_{i} Y^{ij}_{\nu}{\nu}_iv_2 
-\lambda^jv_1v_2+\sum_{i,k} \kappa^{ijk}{\nu^c_i}{\nu^c_k}|^2 \nonumber \\
&+& |\sum_{i}\lambda^i{\nu^c_i}v_2|^2 + \sum_{i}|\sum_{j}Y^{ij}_{\nu}v_2
{\nu^c_j}|^2 +(\frac{g_1^2+g_2^2}{8}) [\sum_{i}|{\nu}_i|^2+|v_1|^2 - 
|v_2|^2]^2 \nonumber \\ 
&+& m_{H_2}^2|v_2|^2 + m_{H_1}^2|v_1|^2 + \sum_{i,j}
(m_{\tilde{L}}^2)^{ij}{\nu}_i^*{\nu}_j +\sum_{i,j}(m_{\tilde{\nu}^c}^2)^{ij}
{\nu^{c^*}_i}{\nu^c_j} \nonumber \\
&+& \left [\sum_{i,j}(A_\nu Y_\nu )^{ij} {\nu}_i{\nu^c_j}v_2 -\sum_{i}
(A_\lambda \lambda )^{i} {\nu^c_i}v_1v_2 + \sum_{i,j,k}\frac{1}{3}(A_\kappa 
\kappa )^{ijk}\nu^c_i\nu^c_j\nu^c_k + {\rm H.c.} \right].\nonumber \\
\label{Vneut}
\eea

One important thing is to notice  that the potential is bounded from below 
because the coefficient of the fourth power of all the eight superfields are 
positive. We shall further assume that all the parameters present in the scalar 
potential are real. From eq.(\ref{Vneut}), the minimization conditions with 
respect to $\nu^c_i~\nu_i,v_2,~v_1$ are 

\beq
2{\sum_{j}} {{u^{ij}_c}} {\zeta^{j}} + \sum_{k} Y^{ki}_{\nu} {r^{k}_c} {v_2^2}
+ {\rho^i \eta} + {\mu} {\lambda^i v_2^2} +\sum_{j} (m^2_{\tilde{\nu}^c})^{ji} 
{\nu^c_j}+(A_x x)^{i}=0,
\label{Minim1}
\eeq
\beq
{\sum_{j}} {Y_{\nu}}^{ij} {v_2} {\zeta^{j}}  +{\gamma_g}{\xi_{\upsilon}}{\nu_i}
+{r^i_c} {\eta} +\sum_{j} (m^2_{\tilde{L}})^{ji} {\nu_j}+\sum_{j} (A_{\nu} 
Y_{\nu})^{ij} {\nu^c_j} v_2 =0,
\label{Minim2}
\eeq
\beq
{\sum_{j}}{\rho^{j}}{\zeta^{j}} + {{\sum_{i}} {r^i_c}^2 v_2}+{\mu^2} v_2
-{\gamma_g}{\xi_{\upsilon}} {v_2} + {\sum_{i}}({A_{\nu}Y_{\nu}})^{ij} 
{\nu_i} {\nu^c_j}-({A_{\lambda} {\lambda}})^i {\nu^c_i} v_1 + {m^2_{H_2}}v_2 
= 0,
\label{Minim3}
\eeq
\beq
-{\sum_{j}}{\lambda^j}v_2 {\zeta^{j}}+{\gamma_g}{\xi_{\upsilon}} {v_1} + 
{\mu^2} v_1 - {\mu} {\sum_{j}} {r^j_c} {\nu_j} + {m^2_{H_1}}v_1-({A_{\lambda} 
{\lambda}})^i {\nu^c_i} v_2 = 0,
\label{Minim4}
\eeq
where
\bea
(A_x x)^{i} &=& \sum_{j} (A_{\nu} Y_{\nu})^{ji} {\nu_j} v_2 + \sum_{j,k} 
({A_\kappa} {\kappa})^{ijk} {\nu^c_j} {\nu^c_k} - (A_{\lambda} {\lambda})^i v_1 v_2,  \nonumber \\
\zeta^{j} &=& \sum_{i,k} 
\kappa^{ijk} {\nu^c_i} {\nu^c_k} + \sum_{k} Y^{kj}_{\nu} v_2 {\nu_k} - 
{\lambda}^j v_1 v_2 , \nonumber \\ 
\eta &=&\sum_{i,j} Y^{ij} _{\nu} {\nu_i} {\nu^c_j} - (\sum_{i} \lambda^i 
{\nu^c_i}) v_1 ,  \nonumber \\
\mu &=& \sum_{i}\lambda^{i} {\nu^c}_{i},  \nonumber \\
\gamma_{g} &=& \frac{1}{4}({g_1^2 + g_2^2}),  \nonumber \\
\xi_{\upsilon} &=& ({\sum_{i} {\nu^2_i} + v_1^2 -v_2^2}),  \nonumber \\
\rho^{i} &=& {\sum_{j}} (Y^{ji}_{\nu} {\nu_j} - \lambda^i v_1),  \nonumber \\
{r^{i}_c} &=& \epsilon^i = {\sum_{j}Y^{ij}_{\nu} {\nu^c_{j}}},  \nonumber \\
{r^{i}} &=& {\sum_{j}Y^{ij}_{\nu} {\nu_{j}}},  \nonumber \\
{u^{ij}_c} &=& {\sum_{k}}\kappa^{ijk}{\nu^c_k}. 
\label{Abbrevations}
\eea
In deriving the above equations, it has been assumed that 
${\kappa}^{ijk}$, $({{A_\kappa} {\kappa}})^{ijk}$,  $Y^{ij}_{\nu}$,
$(A_{\nu} Y_{\nu})^{ij}$, $(m^2_{\tilde{\nu}^c})^{ij}$, 
$(m^2_{\tilde{L}})^{ij}$ are all symmetric in $i,j,k$.

Note that the Dirac masses for neutrinos are given by $m_D^{ij} 
\equiv Y_{\nu}^{ij} v_2$. From present day experiments it is well known that 
neutrino masses are very small. This implies that the neutrino Yukawa 
couplings must also be very small $\sim \cal{O}$ $(10^{-7})$, in order to get 
correct neutrino mass scale using TeV scale seesaw mechanism. This immediately 
tells us that in the limit $Y_{\nu}^{ij} \rightarrow 0$, eq. (\ref{Minim2}) 
implies that $\nu_{i} \rightarrow 0$. So in order to get appropriate neutrino 
mass scale both $Y_{\nu}^{ij}$ and $\nu_{i}$ have to be small.

Ignoring the terms of the second order in $Y_{\nu}^{ij}$ 
and assuming $({\nu^2_i}+v^2_1-v^2_2) \approx (v^2_1-v^2_2)$, 
\textbf{$(m^2_{\tilde{L}})^{ij} = (m^2_{\tilde{L}}) \delta^{ij}$}, we can 
easily solve eq.(\ref{Minim2}) as (using eq. (\ref{Abbrevations}))
\beq
{\nu_i}\approx - \left\{{\frac{{Y_{\nu}}^{ik}{u^{kj}_c} {v_2} - {\mu}{v_1} 
{Y^{ij}_{\nu}}  +(A_{\nu} Y_{\nu})^{ij} v_2}{{\gamma_{g}} ({v_1^2 -v_2^2})+ 
(m^2_{\tilde{L}})}}\right\}{\nu^c_{j}} + 
\left\{{\frac{{Y_{\nu}}^{ij}{\lambda}^j v_1 v^2_2}{{\gamma_{g}} 
({v_1^2 -v_2^2})+ (m^2_{\tilde{L}})}}\right\}.
\label{sneutrino_VEV_simplified}
\eeq
Note from eq.(\ref{sneutrino_VEV_simplified}), that the left handed sneutrinos
can acquire, in general, non-vanishing, non-degenerate VEVs even in the limit 
of zero vacuum expectation values of the gauge singlet sneutrinos. However, 
zero VEVs of all the three gauge singlet sneutrinos is not an acceptable 
solution since in that case no $\mu$-term will be generated. Moreover, one 
needs to ensure that the extremum value of the potential corresponds to the 
minimum of the potential, by studying the second derivatives. 

\section{The scalar sector}
\label{The scalar sector}

The scalar sector of this model enhances from that of MSSM, because of the 
choice of the superpotential in eq.(\ref{superpotential}) (fourth, fifth and 
the sixth term). In this case, the neutral Higgs bosons can mix with both 
the doublet and gauge-singlet sneutrinos. The CP-odd(pseudoscalar) and 
CP-even(scalar) mass squared matrices are now $8\times8$, considering all 
three generations of doublet and singlet sneutrinos. Similarly the charged 
Higgs can mix with the charged sleptons and thus the charged scalar mass 
squared matrix is enhanced to $8\times8$. We have considered only 
the CP-preserving case and hence all the VEVs are chosen to be real. 
The scalar sector of this model has been addressed also in a recent work 
\cite{munoz-lopez-2}. The details of various scalar mass squared matrices are 
given in appendix~{\ref{appenxA}}.



For our analytical and numerical calculations in the later part of the paper, 
we have assumed that $(m_{\tilde{L}}^2)^{ij} = (m_{\tilde{L}}^2) 
\delta^{ij}$, $(m_{\tilde{\nu}^c}^2)^{ij} = (m_{\tilde{\nu}^c}^2) \delta^{ij}$, 
and $Y_\nu^{ij} = 0,\hspace{0.1cm} \text{if} \hspace{0.1cm}i \neq j$. 
We have further assumed that $\kappa^{ijk}$ are flavour-diagonal as well as 
flavour-blind, i.e., $\kappa^{ijk} = \kappa$ if $i=j=k$ and zero otherwise. 
Similarly, we have assumed a flavour-blind coupling $\lambda^i = \lambda$ for 
$i=1,2,3$. We will see that even with such simplifying assumptions, we can fit 
the global three flavour neutrino data in this model. We will use the 
following procedure for all our subsequent analysis. Using the minimization 
conditions, we will solve for the vacuum expectation values $\nu_i$ and 
$\nu^c_i$. We will choose the parameters in such a way that the values of 
$\nu^c_i$ will give an acceptable number for the $\mu$-parameter 
($\mu = \lambda \sum_i \nu^c_i$). As a cross check we confirm the existence of 
two Goldstone bosons in the pseudoscalar and charged scalar mass-squared 
matrices. In addition, we check that all the eigenvalues of the scalar, 
pseudoscalar, and charged scalar mass-squared matrices (apart from the 
Goldstone bosons) should come as positive for a minima. 

Additional constraints on the parameter space can come from the existence 
of false minima. A detailed discussion on this issue has been presented in 
ref. \cite{munoz-lopez-2} and the regions excluded by the existence of false 
minima have been shown. One can check from these figures that mostly the lower 
part of the region allowed by the absence of tachyons, are excluded by the 
existence of false minima. In our analysis, we have chosen the parameter 
points in such a way that they should be well above the regions disallowed by 
the existence of false minima. Nevertheless, in the case of gauge-singlet 
neutrino ($\nu^c$) dominated lightest neutralino (to be discussed later), the value 
of $\kappa$ that we have chosen is 0.07 with two different values of 
$\lambda$, namely, 0.1 and 0.29. In this case, there is a possibility that 
these points might fall into the regions disallowed by the existence of false 
minima. However, we have checked that even if we take the value of $\kappa$ 
to be higher (0.2 or so), with appropriately chosen $\lambda$, our conclusions 
do not change much. For such a point in the parameter space, it is likely that 
the existence of false minima can be avoided. A more detailed study on this 
issue is beyond the scope of the present paper. 

Let us also mention here that the sign of the $\mu$-term is controlled by 
the sign of the VEV $\nu^c$ (assuming a positive $\lambda$), which is 
cotrolled by the signs of $A_\lambda \lambda$ and $A_\kappa \kappa$. 
If $A_\lambda \lambda$ is negative and $A_\kappa \kappa$ is positive then the 
sign of the $\mu$ parameter is negative whereas for opposite signs of the 
above quantities, we get a positive sign for the $\mu$ parameter. 

The scalar mass-squared matrices (both CP odd and CP even) and the vacuum 
expectation values $\nu^c_i$ are not very sensitive to the change in neutrino 
Yukawa couplings ($Y_\nu$ $\sim \cal {O}$ $(10^{-7})$) and the corresponding 
soft parameter $A_\nu Y_\nu$ ($\sim \cal {O}$ $(10^{-4})$ GeV). On the other 
hand, the values of $\tan\beta$ and the coefficients $\lambda$ and $\kappa$ are 
very important in order to satisfy various constraints on the scalar sector 
mentioned earlier. In fig.\ref{scalar_sector_for_tan_beta_values}, we have 
plotted the allowed regions in the ($\lambda$--$\kappa$) plane for 
$\tan\beta=10$. 

\vspace{0.5cm}
\FIGURE{\epsfig{file=Figures/tb10lk.eps,height=5.00cm} 
\caption{Allowed regions in ($\lambda$--$\kappa$) plane which satisfy  
various constraints on the scalar sector, for tan$\beta =10$. $\lambda$ and 
$\kappa$ were allowed to vary from 0.005 to 0.50 and 0.005 to 0.70, respectively.}
\label{scalar_sector_for_tan_beta_values}}

The values of other parameters are chosen to be  
$m_{\tilde{L}} = 400 ~{\text{GeV}}$, $m_{\tilde{\nu}^c} =300 
~{\text{GeV}}, $ $Y_\nu^{11} =5.0\times10^{-7} $,  $Y_\nu^{22} = 
4.0\times10^{-7}$, $Y_\nu^{33} = 3.0\times10^{-7}$, 
$(A_\nu Y_\nu)^{ij} = 1~{\rm TeV}\times Y_\nu^{ij}$,   
$(A_\lambda \lambda) = -1~{\rm TeV}\times \lambda$, and  
$(A_\kappa \kappa) = 1~{\rm TeV}\times \kappa$. The upper limit of the value
of $\kappa$ is taken to be $\sim$ 0.7 because of the constraints coming from 
the existence of Landau pole \cite{munoz-lopez-2}.  
With these values of different parameters satisfying the constraints in the 
scalar sector, we will go on to calculate the neutrino masses and the mixing 
patterns as well as the decays of the lightest neutralino in this model as 
discussed in the next few sections.

It should be mentioned at this point that the radiative corrections to the 
light Higgs mass, can be significant in some regions of the parameter space
as discussed in ref.\cite{munoz-lopez-2}. It has been shown that the light 
Higgs mass larger than the LEP lower limit of 114 GeV can be obtained with the
value of $A_t$ (trilinear coupling in the scalar sector for the stop) within
1-2.4 TeV and when the mixing of the light Higgs with the right-handed 
sneutrino is small. The latter requirement is fulfilled in most of the cases
that we have considered and in some cases the mixing is slightly larger.
However, there is always the freedom of choosing the value of $A_t$
appropriately. Hence, it would be fair to say that the experimental limits on
the light Higgs boson mass can be satisfied in our analysis. The parameter
points we have chosen here are sample points with different dominant
composition of the lightest neutralino. Since we have a large parameter space,
it is always possible to choose a different parameter point with the same
characteristic features satisfying all the experimental constraints. 

\section{The fermionic sector}
\label{The fermionic sector}

\subsection{The neutral fermions}
In this model, because of the breaking of R-parity, two neutral gauginos, 
$\tilde B^0 (=-i{\tilde \lambda}_1)$ and $\tilde W_3^0 
(=-i{\tilde \lambda}_2^3)$ and two neutral higgsinos (${\tilde H}_1^0$ and 
${\tilde H}_2^0$) are now mixed with the neutrinos (both $\nu_i$ and 
$\nu_i^c$). As can be seen from the 
superpotential (fourth and fifth term in eq. $(\ref{superpotential})$), the 
fermionic partners of ${\tilde \nu}^c_i$ and $\tilde{\nu_i}$ mix with the 
neutral higgsinos 
\cite{munoz-lopez_fogliani,munoz-lopez-2}. 
The neutral gauginos are mixed with the left-handed neutrinos through the 
vacuum expectation values of the doublet sneutrinos. Mass matrices for the 
neutral and charged fermion sectors, involving all three generations of 
neutrinos (both doublet and singlet) and charged leptons, have been addressed 
also in \cite{munoz-lopez-2}.

In the weak interaction basis defined by
\bea
{\Psi^0}^T = \left(\tilde B^0, \tilde W_3^0, \tilde H_1^0, 
\tilde H_2^0,{\nu^c_e},{\nu^c_{\mu}},{\nu^c_{\tau}},{\nu_e},{\nu_{\mu}},
{\nu_{\tau}} \right),
\label{neutralino_basis}
\eea
the neutral fermion mass term in the Lagrangian is of the form
\bea
{\mathcal{L}_{neutral}^{mass}} = -\frac{1}{2}{{\Psi^0}^T} \mathcal{M}_n 
{\Psi^0} + \text{H.c.},
\eea
where $\mathcal{M}_n$ includes all three generations of doublet and gauge 
singlet neutrinos and thus it is a $10\times10$ matrix. The massless neutrinos 
become massive due to this mixing with the neutralinos and the gauge singlet 
neutrinos. The three lightest eigenvalues of this $10\times10$ neutralino mass 
matrix correspond to the three light physical neutrinos and their masses have 
to be very small in order to satisfy the experimental data on massive 
neutrinos. The matrix $\mathcal{M}_n$ can be written in the following manner 
\beq
\mathcal{M}_n =
\left(\begin{array}{cc}
M_{7\times 7} & m_{3\times 7}^T \\
m_{3\times 7} & 0_{3\times 3}
\end{array}\right),
\label{neutralino-seesaw}
\eeq
where
\beq
M_{7\times7} =
\left(\begin{array}{ccccccc}
M_1 & 0 & -\frac{g_1}{\sqrt{2}}v_1 & \frac{g_1}{\sqrt{2}}v_2 & 0 & 0 & 0 \\ \\
0 & M_2 & \frac{g_2}{\sqrt{2}}v_1 & -\frac{g_2}{\sqrt{2}}v_2 & 0 & 0 & 0 \\ \\
-\frac{g_1}{\sqrt{2}}v_1 & \frac{g_2}{\sqrt{2}}v_1 & 0 & -{\mu} & 
-{\lambda^e}v_2 & -{\lambda^{\mu}}v_2 & -{\lambda^{\tau}}v_2 \\ \\
\frac{g_1}{\sqrt{2}}v_2 & -\frac{g_2}{\sqrt{2}}v_2 & -{\mu} & 0 & {\rho^e} 
& {\rho^{\mu}} & {\rho^{\tau}}\\ \\
0 & 0 & -{\lambda^e}v_2 & {\rho^e} & 2 {u^{ee}_c} & 2 {u^{e{\mu}}_c} & 
2 {u^{e{\tau}}_c}\\ \\
0 & 0 & -{\lambda^{\mu}}v_2 & {\rho^{\mu}} & 2{u^{{\mu}e}_c} & 
2 {u^{{\mu}{\mu}}_c} & 2 {u^{{\mu}{\tau}}_c}\\ \\
0 & 0 & -{\lambda^{\tau}}v_2 & {\rho^{\tau}} & 2 {u^{{\tau}e}_c} & 
2{u^{{\tau}{\mu}}_c} & 2 {u^{{\tau}{\tau}}_c}
\end{array}\right),
\label{neutralino_7x7}
\eeq
and
\beq
m_{3\times7} =
\left(\begin{array}{ccccccc}
-\frac{g_1}{\sqrt{2}}{\nu_e} & \frac{g_2}{\sqrt{2}}{\nu_e} & 0 & {r^e_c} & 
Y_{\nu}^{ee} v_2 & Y_{\nu}^{e{\mu}} v_2 & Y_{\nu}^{e{\tau}} v_2\\ \\
-\frac{g_1}{\sqrt{2}}{\nu_{\mu}} & \frac{g_2}{\sqrt{2}}{\nu_{\mu}} & 0 & 
{r^{\mu}_c} & Y_{\nu}^{{\mu}e} v_2 & Y_{\nu}^{{\mu}{\mu}} v_2 & 
Y_{\nu}^{{\mu}{\tau}} v_2\\ \\
-\frac{g_1}{\sqrt{2}}{\nu_{\tau}} & \frac{g_2}{\sqrt{2}}{\nu_{\tau}} & 0 &
{r^{\tau}_c}  & Y_{\nu}^{{\tau}e} v_2 & Y_{\nu}^{{\tau}{\mu}} v_2 & 
Y_{\nu}^{{\tau}{\tau}} v_2
\end{array}\right).
\label{neutralino_3x7}
\eeq

Note that the top-left $4\times4$ block of the matrix $M_{7\times7}$ is the 
usual neutralino mass matrix of the MSSM. The bottom right $3\times3$ block is
the Majorana mass matrix of gauge singlet neutrinos, which will be taken as 
diagonal in our subsequent analysis. The entries of $M_{7\times7}$ are in
general of the order of the electroweak scale and the entries of 
$m_{3\times7}$ are much smaller ($\sim$ $\cal O$($10^{-5}$ GeV)). Hence, the 
matrix (\ref{neutralino-seesaw}) has a seesaw structure that will give 
rise to three very light eigenvalues corresponding to three light neutrinos. 
The correct neutrino mass scale of $\sim$ $10^{-2}$ eV can easily be obtained
with such a structure of the $10\times10$ neutralino mass matrix. In this
work our focus would be to see if one can obtain the correct mass-squared
differences and the mixing pattern for the light neutrinos even if we consider
flavour diagonal neutrino Yukawa couplings in eq.(\ref{neutralino_3x7}) (i.e.
with a diagonal Dirac neutrino mass matrix). This makes the analysis simpler 
with a reduced number of parameters and makes the model more predictive. As 
we will show later, it is possible to find out the correct mixing pattern and 
the mass hierarchies (both normal and inverted) among the light neutrinos in 
such a situation, at the tree level. 

In order to obtain the physical neutralino states, one needs to diagonalize 
the $10\times10$ matrix $\mathcal{M}_n$. As in the case of MSSM, the symmetric 
mass matrix $\mathcal{M}_n$ can be diagonalized with one unitary matrix $N$. 
The mass eigenstates are defined by
\beq\label{neutralino_mass_eigenstate}
{\tilde \chi}^0_i= N_{ij} \Psi^0_j, \quad i,j=1,...,10,
\eeq
where the $10\times10$ unitary matrix $N$ satisfies
\beq\label{neutralino_mass_eigenstate_matrixform}
N^* \mathcal{M}_n N^{-1} = \mathcal{M}^0_D,
\eeq
with the diagonal neutralino mass matrix denoted as $\mathcal{M}^0_D$. 
The matrix $N$ may be chosen in such a way that elements of $\mathcal{M}^0_D$ 
are real and non-negative. In our analysis we will assume that all the entries 
in the $10\times10$ neutralino mass matrix $M_n$ are real. Seven eigenstates
of this matrix are heavy, i.e. of the order of the electroweak scale. Out of 
these seven states, there are four states which are usually very similar to 
the MSSM neutralinos. The remaining three states are mostly dominated by 
$\nu^c_i$. It is, in general, very difficult to predict the nature of the 
lightest of these seven states since that depends on several unknown 
parameters. In our analysis of the decays of the lightest supersymmetric 
particle (LSP), we will concentrate on three different possibilities$\colon$
(i) lightest state is dominated by the bino component, (ii) higgsino dominated
lightest state, and (iii) gauge singlet neutrinos $\nu^c_i$ form the 
lightest state. The last possibility is very interesting since in this case we
have the opportunity to produce the gauge singlet neutrinos at the LHC and 
study their properties through the R-parity violating decay modes. This way one
has a direct probe to the seesaw scale at the LHC. 


\subsection{The charged fermions}
In the charged fermion sector, the charged gauginos and charged higgsinos mix 
with the charged leptons because of the presence of the effective bilinear RPV 
parameters $\epsilon^i \equiv \sum_j Y^{ij}_\nu \nu^c_j$ and the sneutrino
VEVs $\nu_i$. This is similar to the case of MSSM with bilinear RPV with the
parameter $\mu$ defined as $\mu = \sum_i \lambda^i \nu^c_i$. Since we want to
calculate the decays of the lightest neutralino, we also need to know the
mass eigenvalues and the mixing matrices in the charged fermion sector. 
Because of this reason, we discuss the chargino mass matrix in some details.    
In the weak interaction basis defined by
\bea
{\Psi^{+T}} = (-i \tilde {\lambda}^{+}_{2}, \tilde{H}_2^{+}, e_{R}^{+}, 
\mu_{R}^{+}, \tau_{R}^{+}), \nonumber \\
{\Psi^{-T}} = (-i \tilde {\lambda}_{2}^{-}, \tilde{H}_1^{-}, e_{L}^{-}, 
\mu_{L}^{-}, \tau_{L}^{-}). \nonumber \\
\label{chargino_basis}
\eea
The charged fermion mass term in the Lagrangian is of the form
\beq\label{chargino_mass_Lagrangian}
{\mathcal{L}_{charged}^{mass}} = -\frac{1}{2} 
\left(\begin{array}{cc}
\Psi^{+^T} & \Psi^{-^T}
\end{array}\right)
\left(\begin{array}{cc}
0_{5\times5} & m_{5\times5}^T \\ \\
m_{5\times5} & 0_{5\times5}
\end{array}\right)
\left(\begin{array}{c}
\Psi^+ \\ \\
\Psi^-
\end{array}\right).
\eeq
Here we have included all three generations of charged leptons and assumed 
that the charged lepton Yukawa couplings are in the diagonal form. Also, 
$-i\lambda^\pm$ are the two-component charged Wino fields and ${\tilde H}_1^-$ 
and ${\tilde H}_2^+$ are the two-component charged higgsino fields. The matrix 
$m_{5\times5}$ is given by
\beq\label{chargino_mass_matrix}
m_{5\times5} =
\left(\begin{array}{ccccc}
M_2 & {g_2}{v_2} & 0 & 0 & 0 \\ \\
{g_2}{v_1} & {\mu} & -{Y_{e}^{ee}}{\nu_e} & -{Y_{e}^{{\mu}{\mu}}}{\nu_{\mu}} &
-{Y_{e}^{{\tau}{\tau}}}{\nu_{\tau}} \\ \\
{g_2}{\nu_e} & -{r^e_c} & {Y_{e}^{ee}}{v_1} & 0 & 0 \\ \\
{g_2}{\nu_{\mu}} & -{r^{\mu}_c} & 0 & {Y_{e}^{{\mu}{\mu}}}{v_1} & 0 \\ \\
{g_2}{\nu_{\tau}} & -{r^{\tau}_c} & 0 & 0 & {Y_{e}^{{\tau}{\tau}}}{v_1}
\end{array}\right).
\eeq
The charged fermion masses are obtained by applying a bi-unitary 
transformation such that 
\beq\label{chargino_mass_eigenstate_matrixform}
U^* m_{5\times5} V^{-1} = \mathcal{M}^{\pm}_D,
\eeq
where $U^*$ and $V$ are two unitary matrices and $\mathcal{M}^{\pm}_D$ is the
diagonal matrix with non-negative entries corresponding to the
physical fermion masses. The two-component mass eigenstates are defined 
by
\bea\label{chargino_mass_eigenstate}
& &\chi^+_i= V_{ij} \Psi^+_j, \nonumber \\
& &\chi^-_i= U_{ij} \Psi^-_j, \quad i,j=1,...,5.
\eea
Nevertheless, we notice that the 13, 14, and 15 elements of the chargino mass 
matrix (eq. (\ref{chargino_mass_matrix})) are vanishing and given the orders of 
magnitude of various parameters, we also see that the values of the other 
off-diagonal entries (except for 12 and 21 elements) are very small. This 
indicates that the physical charged lepton eigenstates will have a very small 
admixture of charged higgsino and charged gaugino states. So we can very well 
assume (also verified numerically) that this mixing has very little effect on 
the mass eigenstates of the charged leptons. Thus, while writing down the 
neutrino mixing matrix, it will be justified to assume that one is  working in 
the basis where the charged lepton mass matrix is already in the diagonal form.


\section{Neutrinos}
\subsection{Seesaw masses} 
When lepton-number violation is allowed, the effective light neutrino mass 
matrix arising via the seesaw mechanism is in general given by

\beq 
{M^{\nu}} = -{m_{3\times7}} {M_{7\times7}^{-1}}
{m_{3\times7}^T}.
\label{seesaw_formula}
\eeq

As discussed in the previous section, ${m_{3\times7}}$ is the so-called Dirac 
neutrino mass matrix and ${M_{7\times7}}$ is the matrix for the heavy states 
and contains $\Delta{L}=2$ mass terms for right chiral neutrinos. In order to 
find out the neutrino mass-squared differences and mixing angles, one must
diagonalize the matrix $M^\nu$ to find out the eigenvalues
and the eigenvectors. Before discussing the detailed numerical results, let us
try to understand the characteristic features of this neutrino mass matrix 
analytically. Let us note that the neutrino mass-squared differences indicate
three possible scenarios for the light neutrino mass spectrum. They are (i)
Normal hierarchy corresponding to $m_1 \approx m_2 \sim \sqrt{\Delta m^2_{21}}$
,~$m_3 \sim \sqrt{|\Delta m^2_{31}|}$, (ii) Inverted hierarchy$\colon$ $m_1 
\approx m_2 \sim \sqrt{|\Delta m^2_{31}|}$, $m_3 \ll 
\sqrt{|\Delta m^2_{31}|}$ and (iii) Degenerate masses$\colon$ $m_1 
\approx m_2 \approx m_3 \gg \sqrt{|\Delta m^2_{31}|}$, where $m_1$, $m_2$,
and $m_3$ are the three light neutrino mass eigenvalues.

\subsubsection{Analytical results}
\label{Analytical results}

Let us make a few simplifications in order to diagonalize the effective
light neutrino mass matrix. One should note that the neutrino mass matrix 
involves the vacuum expectation values for the doublet and the gauge singlet 
sneutrinos. Hence we also make some simplifying assumptions for the parameters
appearing in the scalar sector. Some of these assumptions have already been 
mentioned in the scalar sector but here we repeat them for the convenience 
of the reader. We have defined
\bea
& &{\kappa^{ijk}} = {\kappa}, {\hspace{0.2cm}}\text{if} 
{\hspace{0.2cm}}{i=j=k}, {\hspace{0.2cm}}\text{and zero otherwise}, \nonumber \\
& &{(A_{\kappa} \kappa)^{ijk}} = (A_\kappa \kappa), {\hspace{0.2cm}}\text{if} 
{\hspace{0.2cm}}{i=j=k}, {\hspace{0.2cm}}\text{and zero otherwise}, \nonumber \\
& &{Y_{\nu}^{ij}} = 0,\hspace{0.2cm} \text{if} \hspace{0.2cm} {i\neq j}, 
\nonumber \\
& &(A_\nu Y_{\nu})^{ij} = 0,\hspace{0.2cm} \text{if} \hspace{0.2cm} {i\neq j},
\nonumber \\
& &{\lambda^1}={\lambda^2}={\lambda^3}={\lambda}, \nonumber \\
& &{(A_\lambda \lambda)^1}={(A_\lambda \lambda)^2}={(A_\lambda \lambda)^3}=
{(A_\lambda \lambda)}, \nonumber \\
& & (m^2_{\tilde{L}})^{ij}=(m^2_{\tilde{L}})\delta^{ij}, \nonumber \\
& & (m^2_{\tilde{\nu}^c})^{ij}=(m^2_{\tilde{\nu}^c})\delta^{ij},
\label{assumptions}
\eea
where $i,j,k = e, \mu, \tau$ in the flavour basis. 

With these assumptions one can solve for $\nu^c_i$ from the minimization 
equations for the gauge singlet sneutrinos (eq.(\ref{Minim1})) and the result
is ${\nu^c_1}={\nu^c_2}={\nu^c_3}={\nu^c}$. This can be understood if we 
neglect the terms proportional to ${Y_{\nu}}^2$, ${Y_{\nu}}{\nu}$,
$({A_{\nu}}{Y_{\nu}}){\nu}$ in the minimization equations and assume that
$\lambda, \kappa$ are $\sim {\cal O}$(1) couplings. 

Now let us look at the effective left chiral neutrino mass matrix 
(eq.(\ref{seesaw_formula})) in a little more details. Because of the smallness 
of $Y_\nu$ and $\nu_i$, one can neglect terms containing 
${Y_{\nu}}^2 {\nu^2}$ and  ${Y_{\nu}}^3 {\nu}$. This way one obtains an 
approximate analytical expression for the $3\times3$ neutrino mass matrix.
\bea
& &{M^{\nu}} = 
{\frac{2}{3}}{\frac{A {\nu^c}}{\Delta}}\left(\begin{array}{ccc}
b^2_e & b_eb_\mu & b_e b_\tau  \\ \\
b_eb_\mu & b^2_\mu  & b_\mu b_\tau\\ \\
b_e b_\tau& b_\mu b_\tau & b^2_\tau
\end{array}\right) + {{\frac{1}{6 {\kappa}{\nu^c}}}}\left(\begin{array}{ccc}
-2 a^2_e & a_ea_\mu & a_e a_\tau  \\ \\
a_ea_\mu & -2 a^2_\mu  & a_\mu a_\tau\\ \\
a_e a_\tau & a_\mu a_\tau & -2 a^2_\tau
\end{array}\right) \nonumber \\
& &-{\frac{2{\lambda}{\mu}}{3 \Delta}}{\sum_{i}} Y_{\nu}^{ii} {\nu_i} 
\left(v_2 - {\frac{{2 \lambda}AB}{\Delta}}\right)\left(\begin{array}{ccc}
c^2_e & c_ec_\mu & c_e c_\tau  \\ \\
c_ec_\mu & c^2_\mu  & c_\mu c_\tau\\ \\
c_e c_\tau & c_\mu c_\tau & c^2_\tau
\end{array}\right),\nonumber \\
\label{neutrino_effective}
\eea
where
\bea
\Delta &=& {\lambda}^2 (v^2_1 +v^2_2)^2 + 4 {\lambda} {\kappa} {\nu^c}^2 
v_1 v_2 - 4 M \lambda A{\mu}, ~\mu = 3 \lambda \nu^c,
\nonumber \\
A &=& ({\kappa}{\nu^c}^2 + {\lambda} v_1 v_2), \nonumber \\
\frac{1}{M} &=& \frac{g^2_1}{M_1} +\frac{g^2_2}{M_2}, \\
B &=& v_1(v^2_1 +v^2_2)- 2 M {\mu} v_2, \nonumber \\
a_i &=& Y_{\nu}^{ii} v_2, ~b_i = (Y_{\nu}^{ii} v_1 + 3 {\lambda} {\nu_i}), 
~c_i = {\nu_i}, \nonumber 
\label{clarifications}
\eea
with ${i,j,k} = {e,\mu,\tau}$.

One can rewrite eq. (\ref{neutrino_effective}) in a compact form as follows
\beq
M^{\nu}_{ij} = {\frac{2 A {\nu^c}}{3 \Delta}} {b_{i}} {b_{j}} 
+ {\frac{1}{6 \kappa {\nu^c}}} {a_{i}} {a_{j}} 
(1-3\delta_{ij}) - {\frac{2{\lambda}{\mu}}{3 \Delta}}{\sum_{k}} Y_{\nu}^{kk} {\nu_k} 
\left(v_2 - {\frac{{2 \lambda}AB}{\Delta}}\right) {c_i} {c_j}.
\label{mnuij-compact1}
\eeq

Let us note that the smallness of the left chiral sneutrino VEVs 
(${\nu_i} << v_1, v_2$) allows us to use $m^2_Z \approx {\frac{1}{2}}
(g^2_1+g^2_2) (v^2_1+v^2_2)$ and $\tan{\beta} \approx {\frac{v_2}{v_1}}$.

The coefficients of the first term in eq.(\ref{neutrino_effective}) (or
in eq.(\ref{mnuij-compact1})) is of the order of $\frac{1}{\tilde m}$ whereas
the coefficient of the second term is $\lsim$ $\frac{1}{10{\tilde m}}$, where 
$\tilde m$ is the electroweak (or supersymmetry breaking) scale and we have 
assumed that the relevant mass scales are at $\tilde m$ and $\kappa$ is an 
order one coupling. The value of the coupling $\lambda$ (determines the value
of the $\mu$ parameter), which satisfies the neutrino data as well as the 
constraints in the scalar sector, is taken to be of the order of $10^{-1}$. 
On the other hand, the coefficient of the third term is 
$\sim \frac{1}{{\tilde m}} (\frac{\nu_i}{\tilde m})$. Since 
$\frac{\nu_i}{\tilde m} \sim 10^{-6}$--$10^{-7}$, there is an extra 
suppression factor in the elements of third term in 
eq.(\ref{neutrino_effective}), compared to the first two terms. In addition, 
$b^2_i \sim a^2_i \sim c^2_i$ with a slightly larger value of $a^2_i$ and 
$c^2_i$ compared to $b^2_i$ in most cases. Hence, one can neglect the third 
term of eq.(\ref{neutrino_effective}) in comparison to the first two terms. 
However, in our numerical analysis (discussed later) we have kept all the 
terms in eq.(\ref{neutrino_effective}) and checked that the presence of the 
third term changes the result in an insignificant manner. 

Before going on to find out the expressions for the eigenvalues and the 
eigenvectors of the effective neutrino mass matrix 
(eq.(\ref{neutrino_effective})), let us highlight a few limiting cases which
give us some insight regarding the behaviour of the neutrino mass matrix. 
Neglecting the third term one can rewrite eq.(\ref{mnuij-compact1}) in the 
following manner
\bea
M^{\nu}_{ij}\approx {\frac{v^2_2}{6 \kappa {\nu^c}}}{Y^{ii}_{\nu}} 
{Y^{jj}_{\nu}}(1-3 {\delta{ij}}) &-& {\frac{1}{2 M}}\left[{\nu_i}{\nu_j} +{\frac{v_1 {\nu^c} 
(Y^{ii}_{\nu}{\nu_j}+Y^{jj}_{\nu}{\nu_i})}{\mu}
+{\frac{Y^{ii}_{\nu}Y^{jj}_{\nu}v^2_1 {\nu^c}^2}{\mu^2}}}\right]
\nonumber \\
& &{\times} \left[1-{\frac{v^2}{2 M A {\mu}}}\left({\kappa}{\nu^c}^2 
\sin{2\beta} +{\frac{\lambda {v^2}}{2}}\right)\right]^{-1}.
\label{mnuij-compact2}
\eea
Here we have used $v_2 = v \sin{\beta}, v_1 = v \cos{\beta}, \text{and}
~\mu = 3 {\lambda} {\nu^c}$.

In the limit ${\nu^c}\rightarrow \infty$ and $v \rightarrow 0$, 
eq. (\ref{mnuij-compact2}) reduces to
\beq
\label{gauginoseesaw}
M^{\nu}_{ij} \approx -{\frac{{\nu_i}{\nu_j}}{2 M}},
\eeq
which is the first part of the second term of eq.(\ref{mnuij-compact2}). In 
this case the elements of the neutrino mass matrix are bilinears in the 
left-handed sneutrino VEVs and they appear due to a seesaw effect involving 
the gauginos. This is called the ``gaugino seesaw'' effect and neutrino mass 
generation through this effect is a characteristic feature of the bilinear 
$R-$parity violating model. This effect is present in this model because we 
have seen earlier that the effective bilinear R-parity violating terms are 
generated in the scalar potential as well as in the superpotential through the 
vacuum expectation values of the gauge singlet sneutrinos. Note that the
gaugino seesaw effect can generate mass for only one doublet neutrino.

In the limit $M\rightarrow \infty$, eq.(\ref{mnuij-compact2}) reduces to
\beq
\label{ordinaryseesaw}
M^{\nu}_{ij} \approx{\frac{v^2_2}{6 \kappa {\nu^c}}}{Y^{ii}_{\nu}}
{Y^{jj}_{\nu}}(1-3 {\delta{ij}}),
\eeq
which corresponds to the ordinary seesaw effect between the left handed and
gauge singlet neutrinos. Remember that the effective Majorana masses for 
the gauge singlet neutrinos are given by $M_R = 2 \kappa \nu^c$. The ordinary
seesaw effect can generate, in general, masses for all the three neutrinos. 
Thus depending on the magnitudes and the hierarchies of various diagonal
neutrino Yukawa couplings $Y_\nu^{ii}$, one can generate normal or inverted
hierarchy of neutrino masses (combining with the ``gaugino seesaw" effect) 
corresponding to atmospheric and solar mass squared differences, as 
discussed earlier. In this model it is difficult to obtain a degenerate 
neutrino spectrum and we do not consider this possibility in our subsequent 
analysis. 

Now let us try to find out the approximate analytical expressions for the
eigenvalues and eigenvectors of the effective light neutrino mass matrix using
perturbation theory. Neglecting the third term in 
eq.(\ref{neutrino_effective}), the neutrino mass matrix looks like 
\bea
{M^{\nu}} &=&
{\mathcal{B}}\left(\begin{array}{ccc}
b^2_e & b_eb_\mu & b_e b_\tau  \\ \\
b_eb_\mu & b^2_\mu  & b_\mu b_\tau\\ \\
b_e b_\tau& b_\mu b_\tau & b^2_\tau
\end{array}\right) + {\mathcal{A}}\left(\begin{array}{ccc}
-2 a^2_e & a_ea_\mu & a_e a_\tau  \\ \\
a_ea_\mu & -2 a^2_\mu  & a_\mu a_\tau\\ \\
a_e a_\tau & a_\mu a_\tau & -2 a^2_\tau
\end{array}\right), 
\label{eq_final_neutrino_effective}
\eea
where ${\mathcal{A}} = {\frac{1}{6 {\kappa}{\nu^c}}} $ and ${\mathcal{B}}
={\frac{2}{3}}{\frac{A {\nu^c}}{\Delta}}$. As we have argued above, the 
first matrix in  eq.(\ref{eq_final_neutrino_effective}) can be considered as
the unperturbed one and the second matrix can be treated as a perturbation over
the first one because of the presence of the smaller coefficient $\mathcal{A}$. 
The eigenvalues of the unperturbed matrix are $(0, 0, 
{\mathcal{B}}{(b^2_e + b^2_\mu + b^2_\tau)})$ and the corresponding 
eigenvectors are 
$\left(\begin{array}{ccc}
-\frac{b_{\tau}}{b_e} & 0 & 1
\end{array}\right)^{T}, 
\left(\begin{array}{ccc}
-\frac{b_{\mu}}{b_e} & 1 & 0
\end{array}\right)^{T},  
\left(\begin{array}{ccc}
\frac{b_e}{b_{\tau}} & \frac{b_{\mu}}{b_{\tau}} & 1
\end{array}\right)^{T}$.
With the order of magnitudes of various parameters discussed above, the 
only non-zero eigenvalue determines the atmospheric neutrino mass scale 
corresponding to the normal hierarchical mass pattern for neutrinos. In order
to generate the solar neutrino mass scale one must turn on the perturbation. 
In this case one should use the unperturbed eigenvectors to get the corrections 
to the eigenvalues due  to perturbation. However, since two of the eigenvalues 
are zero, one needs to apply the degenerate perturbation theory to evaluate
the correction to the eigenvalues. To do this, first we construct a complete
set of orthonormal eigenvectors using Gram-Schmidt orthogonalization 
procedure. The set of orthonormal eigenvectors obtained in this case are
\bea
& &\text{y}_1 =
{\frac{b_e}{\sqrt{b^2_e + b^2_{\tau}}}}\left(\begin{array}{c}
-\frac{b_{\tau}}{b_e} \\
0 \\
1
\end{array}\right),\nonumber \\
& &\text{y}_2 =
{\frac{\sqrt{b^2_e + b^2_{\tau}}}{{{\Omega}_b}}}
\left(\begin{array}{c}
-\frac{b_e b_{\mu}}{b^2_e + b^2_{\tau}} \\
1 \\
-\frac{b_{\mu} b_{\tau}}{b^2_e + b^2_{\tau}}
\end{array}\right),\nonumber \\
& &\text{y}_3 =
{\frac{b_{\tau}}{{{\Omega}_b}}}\left(\begin{array}{c}
\frac{b_e}{b_{\tau}} \\
\frac{b_{\mu}}{b_{\tau}} \\
1
\end{array}\right),
\label{eigenvectors}
\eea
where
\beq\label{omegab}
{\Omega}_b={\sqrt{b^2_e + b^2_{\mu} + b^2_{\tau}}}.
\eeq

Using degenerate perturbation theory for this set of orthonormal eigenvectors, 
the modified eigenvalues $m^\prime_{\pm}$ and $m^\prime_3$ are obtained as 
\bea
m^\prime_{\pm} &=& -{\frac{\mathcal{A}}{{\Omega}_b^2}} \left\{{\Pi_{ab}} 
\pm {\sqrt{\left[-3 {{\Omega}^2_b}{\left({\Sigma_{ab}}\right)^2}  +
\left({\Pi_{ab}}\right)^2 \right]}} \right\},
\nonumber \\
m^\prime_3 &=& {\mathcal{B}}{{\Omega}^2_b} - 
{\frac{2{\mathcal{A}}}{{\Omega}_b^2}} \left\{(\sum_{i}{a_i}{b_i})^2 - 
3 {\Lambda}_{ab}\right\},
\label{eigenvalues}
\eea
where
\bea
{\Lambda}_{ab}&=&{{\sum_{i < j}} {a_i} {a_j} {b_i} {b_j}},\nonumber \\
{\Pi_{ab}}&=&{\sum_{i < j}}(a_i b_j+ a_j b_i)^2-{{\Lambda}_{ab}},
\nonumber \\
{\Sigma_{ab}}&=&{\sum_{i \neq j \neq k}}{a_i}{a_j}{b_k}.
\eea

As one can see from eq.(\ref{eigenvalues}), the corrections to the eigenvalues
are proportional to the coefficient $\mathcal{A}$ appearing in 
eq.(\ref{eq_final_neutrino_effective}). This is the effect of the ordinary 
seesaw. Let us note in passing that this effect is absent if only one 
generation of left chiral neutrino is considered, whereas for two and three 
generations of left chiral neutrino the ordinary seesaw effect exists. 
This can be understood from the most general calculation involving 
n-generations of left chiral neutrinos, where the coefficients of 
$\mathcal{A}$ pick up an extra factor $(n-1)$.

\subsubsection{Numerical Results}
\label{Numerical results}

In order to get some idea about the numbers of the mass eigenvalues 
and to make comparisons between the full numerical results and the results 
using approximate analytical expression, we look at a sample point in the 
parameter space. As mentioned earlier, we are considering only the normal 
hierarchical pattern of neutrino masses. The set of parameters are
$M_1 = 325~{\rm GeV}, M_2 = 650~{\rm GeV}, \lambda = 0.06, 
\kappa = 0.65, ~A_\lambda \lambda (-~A_\kappa \kappa) = -1 {\rm TeV} \times \lambda (\kappa)  ~{\rm and}~\tan\beta = 10$.

The choices of diagonal neutrino Yukawa couplings ($Y_\nu^{ii}$) and 
corresponding soft parameters $(A_\nu Y_\nu)^{ii}$ are very 
crucial and we take, for this particular calculation, $Y_{\nu}^{ee} = 
4.57\times{10^{-7}}$, $Y_{\nu}^{\mu \mu} = 6.37\times{10^{-7}}$,
$Y_{\nu}^{\tau \tau} = 1.80\times{10^{-7}}$, $(A_\nu Y_\nu)^{ee} = 
1.57 \times 10^{-4}$ GeV,~$(A_\nu Y_\nu)^{\mu \mu} = 4.70 \times 10^{-4}$ GeV,
~$(A_\nu Y_\nu)^{\tau \tau} = 3.95 \times 10^{-4}$ GeV. Soft masses of 
left handed and right handed sleptons are chosen to be 400 GeV and 300 GeV, 
respectively. Later on we will show the allowed regions in the $Y_\nu$ planes 
which satisfy the experimental data on neutrino masses and mixing. For these 
choices of various parameters, the derived left-handed sneutrino VEVs are 
$\nu_e \sim 10^{-5}~{\rm GeV}, \nu_{\mu}= 1.515 \times 10^{-4}~{\rm GeV}, 
\nu_{\tau} = 2.133 \times 10^{-4}~{\rm GeV}$ and right-handed sneutrino VEVs 
are $\nu^c = -588.74~{\rm GeV}$. With this set of values the masses of three 
neutrinos have been found out by direct diagonalization of the matrix obtained 
using (\ref{seesaw_formula}) and also from the approximate analytical 
expression using (\ref{eigenvalues}). It has been observed that even with 
several simple assumptions (eq.(\ref{assumptions})), all three generations of left 
chiral neutrinos acquire non-vanishing, non-degenerate masses at the 
tree-level. The comparison of the results as obtained from 
(\ref{seesaw_formula}) and from (\ref{eigenvalues}) are given 
in Table \ref{table1}. One can see that these values are within the 3$\sigma$ 
limits shown in eqs.(\ref{eq-msq21})--(\ref{eq-msq31}). However, it should be 
mentioned that if $\lambda$, $\kappa$ are much less than 
$\sim \cal {O}$(1) and $Y_\nu$s are much larger 
than the ones we have considered above, the approximate analytical expression 
does not produce the correct results for the eigenvalues and the eigenvectors. 
Obviously, when the neutrino Yukawa couplings are larger one cannot consider
the second term in eq.(\ref{neutrino_effective}) as a perturbation to the 
first term. In our numerical analysis for obtaining the allowed region of 
parameter space which satisfy the neutrino data, we have done a full numerical 
analysis without using the approximate formula. 

\TABLE[t]
{\begin{tabular}{|c|c|c|c|c|c|}
\hline
 & \multicolumn{3}{c|}{$m_\nu$ (eV) ($\times 10^3$)} 
& $\Delta{m^2_{21}}$(eV$^2$) &  $\Delta{m^2_{31}}$(eV$^2$) \\
\cline{2-4}
 & $m_1$ & $m_2$ & $m_3$ & $(\times 10^5)$ 
& $(\times 10^3)$ \\
\hline
\text{eq.(\ref{seesaw_formula})} & 9.970 & 4.169 & 48.23 & 8.203 & 2.307 \\
\hline
\text{eq.(\ref{eigenvalues})} & 9.468  & 4.168 & 47.71 & 7.228  & 2.187 \\
\hline
\end{tabular}
\caption{Absolute values of the neutrino masses and the mass-squared differences for a sample point of the parameter space discussed in the text. Results for full numerical analysis have been obtained using eq.(\ref{seesaw_formula}). Approximate analytical expressions of eq.(\ref{eigenvalues}) have been used for comparison.\label{table1}}}

The numerical values of the solar and atmospheric mass squared differences 
$\Delta{m^2_{21}}$ and $\Delta{m^2_{31}}$ have also been shown in Table 
\ref{table1} and the results show good agreement. The numerical calculations 
have been performed with the help of a code developed by us using 
Mathematica \cite{math-wolfram}. In our numerical calculations, we have
taken for (i) normal hierarchy: ${m_2}|_{max} < 1.0 \times 10^{-11}$ GeV 
and (ii) inverted hierarchy: ${m_3}|_{max} < 1.0 \times 10^{-11}$ GeV.

\subsection{Neutrino mixing}

The left chiral light neutrinos form a $3\times3$ mass matrix in the flavour 
basis. The unitary matrix which diagonalizes this mass matrix can be 
parameterized as follows \cite{neutrino_pmns}, provided that the charged 
lepton mass matrix is already in the diagonal form

\bea
{\mathcal{U}^\nu} &=&
\left(\begin{array}{ccc}
{c_{12}}{c_{13}} & {s_{12}}{c_{13}} & {s_{13}}{e^{-i {\delta}}} \\ \\
-{s_{12}}{c_{23}}-{c_{12}}{s_{23}}{s_{13}}{e^{-i {\delta}}} & {c_{12}}{c_{23}}
-{s_{12}}{s_{23}}{s_{13}}{e^{-i {\delta}}}  & {s_{23}}{c_{13}}\\ \\
{s_{12}}{s_{23}}-{c_{12}}{c_{23}}{s_{13}}{e^{-i {\delta}}} & -{c_{12}}{s_{23}}
-{s_{12}}{c_{23}}{s_{13}}{e^{-i {\delta}}}  & {c_{23}}{c_{13}}
\end{array}\right),
\label{neutrino_mixing_3x3}
\eea

where $c_{ij} = \cos{\theta_{ij}}$,  $s_{ij} = \sin{\theta_{ij}}$, and
$i,j$ runs from 1 to 3. Various neutrino oscillation experiments indicate that
$\theta_{12} \approx 34^\circ$, $\theta_{23} \approx 45^\circ$, and $\theta_{13}
\leq 13^\circ$ \cite{osc_experiments1, osc_experiments2}. This pattern is 
known as bilarge mixing. In order to understand the consequences of such 
mixing in the zeroth order, one can approximately take $\theta_{23} 
= 45^\circ$, $\sin{\theta_{12}} = {\frac{1}{\sqrt{3}}}$ and $\theta_{13} 
\approx 0^\circ$, something known as tribimaximal structure 
\cite{tribimaximal}. Then the unitary matrix turns out to be

\bea
{\mathcal{U}_{3\times3}^{\nu}} &=&
\left(\begin{array}{ccc}
\sqrt{\frac{2}{3}} & {\frac{1}{\sqrt{3}}} & 0\\ \\
-{\frac{1}{\sqrt{6}}} & {\frac{1}{\sqrt{3}}} & {\frac{1}{\sqrt{2}}}\\ \\
{\frac{1}{\sqrt{6}}} & -{\frac{1}{\sqrt{3}}} & {\frac{1}{\sqrt{2}}}
\end{array}\right).
\label{tribimaximal_form}
\eea
Given the three mass eigenvalues $m_1, m_2, m_3$, it is possible to 
use the matrix  ${\mathcal{U}^{\nu}}$ to obtain the mass matrix in 
the flavour basis as follows,

\beq
{\mathcal{U}^{\nu^{-1}}} {M^{\nu}}
{\mathcal{U}^{\nu}} ={M^{\nu}_{diag}},
\eeq
where
\bea
{M^{\nu}_{diag}}&=&
\left(\begin{array}{ccc}
m_{1} & 0 & 0\\ \\
0 & m_{2} & 0\\ \\
0 & 0 & m_{3}
\end{array}\right).
\eea
We will numerically diagonalize the neutrino mass matrix $M_\nu$ obtained 
in eq.(\ref{seesaw_formula}) and also use the approximate analytical method 
to find out the neutrino mixing matrix $U^\nu$. We will also compare the 
results obtained using these two methods. However, when we will scan the 
parameter space to find out the allowed regions where the neutrino 
experimental data are satisfied, we shall use the full numerical procedure. 
The advantage of having the approximate analytical expression is that it can 
give us some insight regarding the conditions on the model parameters for 
which the bilarge mixing is obtained. We can verify these predictions 
numerically in some regions of the parameter space. We will try to find out the
regions in the models parameters where the numbers in 
eqs.(\ref{eq-msq21})--(\ref{eq-mixing-angles}) are reproduced. 

\subsubsection{Analytical results}
\label{Analytical results-2}

One can construct the neutrino mixing matrix analytically  using the degenerate
perturbation theory. With the set of orthonormal eigenvectors in 
eq. (\ref{eigenvectors}) and the eigenvalues in eq.(\ref{eigenvalues}), it is 
possible to write down the eigenvectors of (\ref{eq_final_neutrino_effective}) 
in the following form
\beq
(\mathcal{Y}_1)_{3\times1} = {\alpha_1}{y_1} + {\alpha_2}{y_2},
\label{evec_anal1}
\eeq
\beq
(\mathcal{Y}_2)_{3\times1} = {{\alpha}^\prime}_{1}{y_1} + {{\alpha}^\prime}_{2}
{y_2},
\label{evec_anal2}
\eeq
\beq
(\mathcal{Y}_3)_{3\times1} = {y_3},
\label{evec_anal3}
\eeq
where ${\alpha_1}$, ${\alpha_2}$, $\alpha^\prime_1$, 
$\alpha^\prime_2$, are calculated using degenerate perturbation 
theory and their analytical expressions are given by
\beq
\alpha_1 = \pm \left( \frac {h_{12}} {\sqrt{h^2_{12} + (h_{11} - 
m^\prime_+)^2}}\right),
\label{alph1}
\eeq

\beq
\alpha_2 = \mp \left( \frac {h_{11} - m^\prime_+} {\sqrt{h^2_{12} + (h_{11} 
- m^\prime_+)^2}}\right),
\label{alph2}
\eeq

\beq
\alpha^\prime_1 = \pm \left( \frac {h_{12}} {\sqrt{h^2_{12} + (h_{11} 
- m^\prime_-)^2}}\right),
\label{alph1prm}
\eeq

\beq
\alpha^\prime_2 = \mp \left( \frac {h_{11} - m^\prime_-} 
{\sqrt{h^2_{12} + (h_{11} - m^\prime_-)^2}}\right).
\label{alph2prm}
\eeq
Here $m^\prime_+$, $m^\prime_-$ are given by eq.(\ref{eigenvalues}) and 
$h_{11}$, $h_{12}$ are given by
\beq
h_{11} = -\frac {2{\mathcal{A}} \left({a^2_{\tau}} {b^2_e} + 
{a_e}{a_{\tau}}{b_e}{b_{\tau}} + {a^2_e} {b^2_{\tau}}\right)}{b^2_+},
\label{h11}
\eeq
and
\beq
h_{12} = \frac{{\mathcal{A}}\left[a_{\mu}(a_{\tau} b_e - a_e b_{\tau}){b^2_+} 
- {b_{\mu}}\left(2 b_e b_{\tau}{a^2_-}+ a_e 
a_{\tau}{b^2_-}\right)\right]}{{\Omega_b}{b^2_+}},
\label{h12}
\eeq
where
\bea\label{h12_specifications}
b^2_{\pm} = (b^2_e \pm b^2_{\tau}), \nonumber \\
a^2_- = (a^2_e - a^2_{\tau}),
\eea
and $\Omega_b$ has been defined in eq.({\ref{omegab}}). 

The neutrino mixing matrix $U^\nu$ can be constructed using these eigenvectors
in eqs.(\ref{evec_anal1})-(\ref{evec_anal3}) and it looks like
\bea
{U^{\nu}}&=&
\left(\begin{array}{ccc}
{\mathcal{Y}_1} & {\mathcal{Y}_2} & {\mathcal{Y}_3}
\end{array}\right)_{3\times3}.
\label{mneutrino_mixing_numerical}
\eea
Looking at the expressions for the eigenvectors, one can immediately draw
a few conclusions regarding the behaviours of the neutrino mixing angles with
the model parameters. For example, the (13) mixing angle $\theta_{13}$ is 
given by
\bea
\sin^2\theta_{13} = \frac{b^2_e}{b^2_e + b^2_\mu + b^2_\tau}.
\label{reactor_analytical}
\eea
If we want the (13) mixing angle to be small then one must have 
$b^2_e \ll (b^2_\mu + b^2_\tau)$. On the other hand, the (23) mixing angle
$\theta_{23}$ is given by
\bea
\sin^2\theta_{23} = \frac{b^2_\mu}{b^2_\mu + b^2_\tau}.
\label{atmos_analytical}
\eea
So, if the (23) mixing is maximal then one would expect $b^2_\mu = b^2_\tau$.
The solar mixing angle $\theta_{12}$ is approximately given by
\bea
\sin^2\theta_{12} \approx 1 - {(\alpha^\prime_1 
+ \alpha^\prime_2 {\frac{b_e}{b_\tau}})}^2,
\label{solar_analytical}
\eea
where $\alpha^\prime_1$ and $\alpha^\prime_2$ are given by 
eqs.(\ref{alph1prm}) and (\ref{alph2prm}), respectively. 
In order to have $\theta_{12} \sim 35^\circ$, the square root of the second 
term on the right hand side of eq.(\ref{solar_analytical}) should 
be approximately 0.8. In the next sub-section we discuss the patterns of 
neutrino mixing in this model numerically, and show the allowed regions of 
the parameter space where the neutrino experimental data are satisfied.
\vspace{0.5cm}
\FIGURE{\epsfig{file=Figures/Ssq23nn.eps,height=5.00cm} 
\caption{Scatter plot of the neutrino mixing angle $\sin^2\theta_{23}$ as 
a function of the ratio $\frac{b^2_\mu}{b^2_\tau}$. Values of other parameters
are described in the text. The lightest neutralino (LN) is either bino or 
higgsino dominated.}
\label{sinsqtheta23_bmubtau}}
\vspace{0.5cm}
%

\subsubsection{Numerical Results}
\label{Numerical results-2}

Let us first calculate the neutrino mixing angles for the parameter point 
discussed in Table \ref{table1}. As we have discussed earlier, this parameter 
point generates the normal hierarchical pattern of neutrino masses. In Table 
\ref{table2}, the three mixing angles are shown and they have been evaluated 
using the direct numerical calculation in eq.(\ref{seesaw_formula}) as well 
as using the approximate analytical expressions in 
eq.(\ref{mneutrino_mixing_numerical}). We want to emphasize once
again that the approximate formulae have been used just to get some idea about
the behaviours of the neutrino masses and mixing angles with various model 
parameters. This way we can identify the relevant parameters which crucially
control the neutrino masses and mixing angles in different regions of the 
parameter space. However, these formulae are not valid everywhere in the 
allowed parameter space and in all the plots shown in this paper we have used
full numerical calculation using eq.(\ref{seesaw_formula}). 

\TABLE
{\begin{tabular}{|c|c|c|}
\hline
\text{mixing angles in degree}  & \text{Using (\ref{seesaw_formula})} & 
\text{Using (\ref{mneutrino_mixing_numerical})}\\
\hline
$\theta_{12}$ & 36.438 & 37.287 \\
\hline
$\theta_{13}$ & 9.424 & 6.428 \\
\hline
 $\theta_{23}$& 38.217 & 42.675 \\
\hline
\end{tabular}
\caption{Neutrino mixing angles for a sample parameter point discussed in 
Table \ref{table1}.  Results are shown using eq.(\ref{seesaw_formula}) and
eq.(\ref{mneutrino_mixing_numerical}).\label{table2}}}

We have taken suitable values for $\lambda$ and $\kappa$ in such a way that
they fall in the region allowed by the constraints in the scalar sector 
(similar to the region shown in fig. \ref{scalar_sector_for_tan_beta_values}). 
We can see that for this choice of the parameter space, numerical and 
approximate analytical results give quite good agreement. Naturally, one would 
be interested to check the predictions made in 
eqs. (\ref{reactor_analytical}), (\ref{atmos_analytical}), 
and (\ref{solar_analytical}) over a wide region in the parameter space and see
the deviations from the full numerical calculations. This has been shown in 
fig. \ref{sinsqtheta23_bmubtau}, where we have plotted the value of 
$\sin^2\theta_{23}$ as a function of the ratio ${b^2_\mu}/{b^2_\tau}$. 
%

We can see from this figure that for $b^2_\mu = b^2_\tau$, the value of 
$\sin^2\theta_{23}$ varies in the range 0.41 -- 0.44, which corresponds to 
$\theta_{23}$ between 40$^\circ$ and 42$^\circ$. On the other hand, 
eq.(\ref{atmos_analytical}) tells that for $b^2_\mu = b^2_\tau$, 
$\sin^2\theta_{23} = 0.5.$ So we see that in this case the result from the 
numerical calculation is reasonably close to the prediction from the 
approximate analytical formula. The choices for various parameters are
given below. For the gaugino mass parameters $M_1$ and $M_2$, we take two 
different sets of values which give us either a bino dominated lightest 
neutralino or a higgsino dominated lightest neutralino. In order to have
a bino dominated lightest neutralino, our choices are $M_1 = 110~{\rm GeV}, 
M_2 = 220~{\rm GeV}$, and for a higgsino dominated lightest neutralino we
take $M_1 = 325~{\rm GeV}, M_2 = 650~{\rm GeV}$. The same choices will be 
made for the gaugino mass parameters when we discuss the decays of the lightest
neutralino in Sec. VI. Our choice of the ratio of the gaugino mass 
parameters at the electroweak scale is motivated by the assumption of 
universal gaugino mass at the grand unified theory scale. The value of 
$\kappa$ is taken as 0.65 which satisfies the constraints from the scalar 
sector. We have taken two different values of $\lambda$ corresponding to a 
bino or a higgsino dominated lightest neutralino. For the bino dominated case
$\lambda$ = 0.13, and for the higgsino dominated case $\lambda$ = 0.06. The
corresponding values for $A_\lambda \lambda = -\lambda \times 1 {\rm TeV}$ and 
$A_\kappa \kappa = \kappa \times 1 {\rm TeV}$. The three diagonal neutrino Yukawa 
couplings ($Y_\nu^{ii}$) vary randomly in different ranges 
\bea
3.55 \times 10^{-7} &\leq& Y_\nu^{11} \leq 5.45 \times 10^{-7} \nonumber \\
5.55 \times 10^{-7} &\leq& Y_\nu^{22} \leq 6.65 \times 10^{-7} \nonumber \\
1.45 \times 10^{-7} &\leq& Y_\nu^{33} \leq 3.35 \times 10^{-7}.
\eea
The corresponding soft parameters ($A_\nu^{ii}$) also vary randomly in 
different ranges such that the parameters $(A_\nu Y_\nu)^{ii}$ 
effectively vary as follows
\bea
1.25 \times 10^{-4} &\leq& (A_\nu Y_\nu)^{11} \leq 1.95 \times 10^{-4} \nonumber \\
3.45 \times 10^{-4} &\leq& (A_\nu Y_\nu)^{22} \leq 4.95 \times 10^{-4} \nonumber \\
2.35 \times 10^{-4} &\leq& (A_\nu Y_\nu)^{33} \leq 4.20 \times 10^{-4}.
\eea 
The allowed regions in the $\lambda -~\kappa$ plane are not very sensitive to the values of $Y_\nu$ and $A_\nu Y_\nu$ due to their smallness. Hence we choose them different for different cases, in order to accommodate the three flavour global neutrino data.

\vspace{0.5cm}
\DOUBLEFIGURE{Figures/Ssq13nn.eps,height=5.00cm}{Figures/Ssq12nn.eps,height=5.00cm}{{$\sin^2\theta_{13}$ as a function of the ratio 
$\frac{b^2_e}{b^2_\mu + b^2_\tau}$. Values of other parameters
are the same as in fig.2.}\label{sinsqtheta13_bsqratios}}{{$\sin^2\theta_{12}$ as a function of  
$(\alpha_1^{\prime}+\alpha_2^{\prime}\frac{b_e}{b_\tau})^2$. 
One can see that as 
$(\alpha_1^{\prime}+\alpha_2^{\prime}\frac{b_e}{b_\tau})^2\rightarrow0.50$, 
$\sin^2\theta_{12}$ tends to $0.50$, as predicted by the analytical formula. Values of other parameters are the same as in fig.2.}
\label{sinsqtheta12_alpha1primesq}}

Note that in some parts of these ranges we have considered a bino dominated 
lightest neutralino and in some parts we have taken a higgsino dominated 
lightest neutralino with some overlapping regions. The values of other 
parameters are chosen to be $m_{\tilde{L}} = 400~{\text{GeV}}$, 
$m_{\tilde{\nu}^c} =300~{\text{GeV}}$ and $\tan\beta = 10.$
We have assumed that the phase $\delta$ appearing in the mixing matrix 
(\ref{neutrino_mixing_3x3}) is zero. One important thing to notice is that 
even with a flavour diagonal structure of the neutrino Yukawa couplings 
$Y_\nu$, one can obtain the required two large mixing angles for the 
neutrinos. The variations of other two mixing angles with the relevant 
parameters are shown in Figs. 
\ref{sinsqtheta13_bsqratios} and \ref{sinsqtheta12_alpha1primesq}.

In fig.\ref{normal_hierarchical_scheme_for_bino-dominated_case}, we have shown
the regions in the various $Y_\nu$ planes satisfying the three flavour global 
neutrino data. The values of other parameters are as in 
fig.\ref{sinsqtheta23_bmubtau} for the case where the lightest neutralino is 
bino dominated. We can see from these figures that the allowed values of 
$Y_\nu$s show a mild hierarchy such that $Y_\nu^{22} > Y_\nu^{11} > 
Y_\nu^{33}$. 

Similar studies have been performed for the inverted 
hierarchical case and the allowed region shows that the magnitudes of the 
neutrino Yukawa couplings are larger compared to the case of normal 
hierarchical scheme of the neutrino masses with a different hierarchy among
the $Y_\nu$s themselves ($Y_\nu^{11} > Y_\nu^{22} > Y_\nu^{33}$). In this case
$\sin^2\theta_{12}$ shows an increasing behaviour with the ratio 
$b^2_e/b^2_\mu$, similar to the one shown by $\sin^2\theta_{23}$ with 
$b^2_\mu/b^2_\tau$ in the normal hierarchical scenario. On the other hand,
$\sin^2\theta_{23}$ shows a decreasing behaviour with $b^2_\mu/b^2_\tau$. In 
all these cases, the solar and atmospheric mass-squared differences are within 
the 3$\sigma$ limits. 
\vspace{0.5cm}
\FIGURE{\epsfig{file=Figures/tb10nnBDYnu12.eps,height=5.00cm} 
\vspace{0.75cm}
\epsfig{file=Figures/tb10nnBDYnu13.eps,height=5.00cm} 
\vspace{0.65cm}
\epsfig{file=Figures/tb10nnBDYnu23.eps,height=5.00cm} 
\caption{Plots for normal hierarchical scheme of neutrino mass in 
$Y^{11}_\nu ~-~Y^{22}_\nu$,~$Y^{11}_\nu ~-~Y^{33}_\nu$ and  $Y^{22}_\nu ~-
~Y^{33}_\nu$ plane when the lightest neutralino (LN) is bino dominated.}
\label{normal_hierarchical_scheme_for_bino-dominated_case}}

The case of $\nu^c$ dominated lightest neutralino has also been studied and 
it shows a very interesting and different behaviour compared to the bino and 
higgsino dominated cases. In this case, the dominant contribution in the 
neutrino mass matrix (eq.(\ref{mnuij-compact1})) comes from the term 
proportional to $a_i a_j$. The terms proportional to $b_i b_j$ should be 
considered as a perturbation. Hence, in the normal hierarchical scenario of
neutrino masses, one would expect that $\sin^2\theta_{23}$ is proportional
to $a^2_\mu/a^2_\tau$. This is exactly what we see in 
fig.\ref{fig-nuc-normal-sinsqtheta23}. 

Note that for $a^2_\mu = a^2_\tau$, the mixing becomes maximal. 
On the other hand, the solar mixing angle is controlled mostly by the 
quantity $b^2_e/b^2_\mu$ and shows an increasing behaviour with this ratio.  
In the case of inverted hierarchical scenario of neutrino 
masses, $\sin^2\theta_{23}$ shows a decreasing behaviour with the ratio
$b^2_\mu/b^2_\tau$ whereas $\sin^2\theta_{12}$ shows an increasing pattern 
with $b^2_e/b^2_\mu$. However, we do not show these plots here. 


\section{Decays of the lightest neutralino}
\label{Decay of the lightest neutralino}
Let us now look at some decay processes which can be considered as the typical 
consequence of this model. It is obvious that because of the R-parity 
violation there will be no stable lightest supersymmetric particle (LSP) 
present in this model. Here we consider the case where the lightest neutralino 
(${\tilde \chi}^0_7$ in our notation to be described below) is the LSP 
(or NLSP in some cases) and study its decay pattern in the R-parity
violating channels. In particular, we will consider the case where 
$m_{{\tilde \chi}^0_7} > m_{W^\pm}$, so that the three-body decays are less 
important compared to the two-body decays ${\tilde \chi}^0_7 \rightarrow Z + 
\nu_{e,\mu,\tau}$, ${\tilde \chi}^0_7 \rightarrow W^\pm + e^\mp$, 
${\tilde \chi}^0_7 \rightarrow W^\pm + \mu^\mp$, and 
${\tilde \chi}^0_7 \rightarrow W^\pm + \tau^\mp$. The required Feynman rules 
for the computation of the decay of the lightest neutralino are given in 
the Appendix {\ref{appenxB}}. Let us also remark that the lightest neutralino
LSP can also decay to $h + \nu$, if it is kinematically allowed, where $h$ is
the MSSM-like lightest Higgs boson. However, for our illustration purposes we
have considered the mass of the lightest neutralino in such a way that this 
decay is either kinematically forbidden or very much suppressed (assuming a 
lower bound on the mass of $h$ to be 114 GeV). Even if this decay branching 
ratio is slightly larger, it is usually smaller than the branching ratios in
the ($\ell_i + W$) channel. Hence, this will not affect our conclusions 
regarding the ratios of branching ratios in the charged lepton channel 
($\ell_i + W$), to be discussed later. The lightest neutralino decay 
${\tilde \chi}^0_7 \rightarrow \nu + {\tilde \nu}^c$, where ${\tilde \nu}^c$ 
is the scalar partner of the gauge singlet neutrino $\nu^c$, is always very 
suppressed. We will discuss more on this when we consider a $\nu^c$ dominated 
lightest neutralino.

Consider the following decay process 
\beq\label{general_decay}
\tilde{\chi}_i \longrightarrow \tilde{\chi}_j + V ,
\eeq
where $\tilde{\chi}_{i(j)}$ is either a neutralino or chargino, with  
mass $m_{i(j)}$ and $V$ is the gauge boson which is either $W^{\pm}$ 
or $Z$, with mass $m_v$. The masses $m_i$ and $m_j$ are positive.

The decay width for this process in eq.(\ref{general_decay}) is given 
by \cite{Gunion-Haber-2, Franke-Frass}

\beq\label{general_decay_width_formula}
\Gamma \left(\tilde{\chi}_i \longrightarrow \tilde{\chi}_j + V\right) = 
\frac{g^2 \mathcal{K}^{1/2}}{32 ~\pi m^3_i m^2_W} \times \left\{\left( G^2_L + G^2_R \right)\mathcal{F} - G^*_L G_R 
~\mathcal{G} \right \},\nonumber \\
\eeq
where $\mathcal{F}$, $\mathcal{G}$ are functions of $m_i, m_j, m_v$ and 
given by
\bea\label{general_decay_width_formula_specifications}
& & \mathcal{F}(m_i, m_j, m_v) =  \mathcal{K} + 3 ~m^2_v \left( m^2_i + m^2_j 
- m^2_v \right), \nonumber \\ 
& & \mathcal{G}(m_i, m_j, m_v) =  12 ~\epsilon_i \epsilon_j m_i m_j m^2_v, \nonumber \\
\eea
with $\epsilon_i(j)$ carrying the actual signs $(\pm 1)$ of the 
neutralino masses. The chargino masses must be positive.
The kinematical factor $\mathcal{K}$ is given by
\beq\label{kinematical_factor}
\mathcal{K}(m^2_i, m^2_j, m^2_v) = \left( m^2_i + m^2_j - m^2_v \right)^2 
- 4 ~m^2_i m^2_j.
\eeq
\vspace{0.5cm}
\FIGURE{\epsfig{file=Figures/Ntb10nnNCDSSQ23amat-1.eps,height=5.00cm} 
\caption{Scatter plot of the neutrino mixing angle $\sin^2\theta_{23}$ as 
a function of the ratio $\frac{a^2_\mu}{a^2_\tau}$. Values of other parameters
are described in the text. The lightest neutralino (LN) is $\nu^c$ dominated.}
\label{fig-nuc-normal-sinsqtheta23}}
\vspace{0.5cm}
In order to derive eq.(\ref{general_decay_width_formula}), we have used 
the relation $m^2_W = m^2_Z \cos^2\theta_W$ and since $\langle \tilde{\nu}_i 
\rangle << v_1, v_2$, some of the MSSM relations still hold good.
The factors $G_L$, $G_R$ are given here for some possible decay modes
\bea\label{some_possible_decay_modes}
& &\text{For} \quad \tilde{\chi}^0_i \longrightarrow \tilde{\chi}^0_j Z 
\nonumber \\
& & G_L = O^{\prime \prime L}_{ji}, \quad  G_R = O^{\prime \prime R}_{ji},
\nonumber \\
\quad \
& &\text{For} \quad \tilde{\chi}^0_i \longrightarrow \tilde{\chi}^+_j {W^-} 
\nonumber \\
& & G_L = O^{L}_{ij}, \quad  G_R = O^{R}_{ij},\nonumber \\
\eea
where $O^{L(R)}_{ij}$ and $O^{\prime \prime L(R)}_{ji}$ are given by 
eq.(\ref{specifications_of_symbols_used_in_mass_eigenstates_of_CN_and_NN}) and 
eq.(\ref{specifications_of_symbols_used_in_mass_eigenstates_of_CN_and_NN2}).

Now consider the following decays
\bea\label{two-selected-decays}
& &\tilde{\chi}^0_{LN} \longrightarrow Z + \nu, \nonumber \\
& &\tilde{\chi}^0_{LN} \longrightarrow W^{\pm} + \ell^{\mp}, \nonumber \\
\eea
where $\tilde{\chi}^0_{LN}$ stands for lightest neutralino and $\ell= e, \mu,
\tau$.
At this stage let us discuss our notation and convention for these decays. The 
neutralino mass matrix is  a 10$\times$10 mass matrix which includes the left 
handed as well as the gauge-singlet neutrinos. If the mass eigenvalues of this 
matrix are arranged in the descending order then the three lightest eigenvalues of 
this 10$\times$10 neutralino mass matrix would correspond to the three light 
neutrinos. Out of the remaining seven heavy eigenvalues, the lightest one 
is denoted as the lightest neutralino. Thus, in our notation 
$\tilde{\chi}^0_7$ is the lightest neutralino and 
$\tilde{\chi}^0_{j+7}, \text{where}$ $j=1,2,3$ correspond to the three light 
neutrinos. Similarly, for the chargino masses, $\tilde{\chi}^{\pm}_{l+2}$ 
($l=1,2,3$) corresponds to the charged leptons $e, \mu, \tau$.

So for $\tilde{\chi}^0_{LN} \rightarrow Z + \nu$, which is also 
$\tilde{\chi}^0_7 \rightarrow Z + \tilde{\chi}^0_{j+7}$ ($j = 1,~2,~3 $), 
one gets from eq.(\ref{some_possible_decay_modes}) and
eq.(\ref{specifications_of_symbols_used_in_mass_eigenstates_of_CN_and_NN})
\bea\label{G_L-&-G_R-for-first-of-two-selected-decays}
& & G_L = - \frac{1}{2} N_{j+7,3} N^*_{73} + \frac{1}{2} N_{j+7,4} N^*_{74} 
- \frac{1}{2} N_{j+7,k+7} N^*_{7,k+7}, \nonumber \\
& & G_R = -G^*_L,\nonumber \\
\eea
where $~j,k = 1,~2,~3$ and this in turn modifies 
eq.(\ref{general_decay_width_formula}) as 
\beq\label{decay-width-for-process-1}
\Gamma \left(\tilde{\chi}^0_7 \rightarrow Z + ~\tilde{\chi}^0_{j+7}
\right) = \frac{g^2 \mathcal{K}^{1/2}}{32 ~\pi  m^3_{{\tilde \chi}^0_7} 
m^2_W} \times \left\{ 2 ~G^2_L \mathcal{F} + G^{*^2}_L ~\mathcal{G} \right \},
\eeq
with $m_i = m_{{\tilde \chi}^0_7}$, $m_j = m_{\nu} \approx 0$ and $m_v = m_Z$.

Let us now consider the other decay which is $\tilde{\chi}^0_{LN} \rightarrow 
W^{\pm} + \ell^{\mp}$ or equivalently $\tilde{\chi}^0_7 \rightarrow 
W^{\pm} + \tilde{\chi}^{\mp}_j (j=3,4,5)$.

For the process $\tilde{\chi}^0_7 \rightarrow W^{-} + \tilde{\chi}^{+}_j$
\bea\label{decay-width-for-process-2-a}
& &\Gamma \left(\tilde{\chi}^0_7 \rightarrow W^- 
+ ~\tilde{\chi}^+_j\right) = \frac{g^2 \mathcal{K}^{1/2}}{32 ~\pi 
m^3_{{\tilde \chi}^0_7} m^2_W} \times \left\{ \left(G^2_L + G^2_R \right) \mathcal{F} -G^*_L G_R 
~\mathcal{G} \right \},\nonumber \\
& & G_L = N_{72} V^*_{j1} - \frac{1}{\sqrt{2}} N_{74} V^*_{j2},\nonumber \\
& & G_R = N^*_{72} U_{j1} + \frac{1}{\sqrt{2}} N^*_{73} U_{j2} 
+ \frac{1}{\sqrt{2}} N^*_{7,k+7} U_{j,k+2},\nonumber \\
& & (k = 1,~2,~3),
\eea
where eq.(\ref{some_possible_decay_modes}) and
eq.(\ref{specifications_of_symbols_used_in_mass_eigenstates_of_CN_and_NN2}) has been used. The process $\tilde{\chi}^0_7 \longrightarrow W^{+} 
+ \tilde{\chi}^{-}_j$ is obtained by charge conjugation of the process
in eq.(\ref{decay-width-for-process-2-a}).

\subsection{Correlation between the lightest neutralino decays and neutrino 
mixing angles}

Correlations between the lightest neutralino decays and neutrino mixing angles
will depend on the nature of the lightest neutralino as well as on the 
mass hierarchies of the neutrinos, i.e.  whether we have a normal 
hierarchical pattern of neutrino masses or an inverted one. 
In this section we look into these possibilities in details and consider three 
different cases for the dominant component of the lightest neutralino. We 
consider that the lightest neutralino is (1) bino dominated, 
(2) higgsino dominated, and (3) $\nu^c$ dominated. For each of these 
cases we consider both the normal and the inverted hierarchical pattern 
of neutrino masses. We show that in these different cases, the 
ratio of branching ratios of certain decays of the lightest neutralino 
correlates with the neutrino mixing angles. In some cases
the correlation is with the atmospheric and the reactor angle and in other 
cases the ratio of the branching ratios correlates with the solar mixing
angle and in some cases there is no correlations at all. Let us now 
study these possibilities case by case. The interesting 
difference between this study and similar studies with bilinear R-parity 
violating scenario \cite{correlations_neutralino} in the MSSM is the 
presence of a gauge singlet neutrino dominated lightest neutralino. We 
will see later that in this case the results can be very different from 
the bino or higgsino dominated lightest neutralino. The lightest neutralino
decays in neutrino mass models with spontaneous R-parity violation have been
studied in ref.\cite{spontaneous-rpv-neutralino}.   

\subsubsection{Bino dominated lightest neutralino}
We will assume that the gaugino masses are unified at the grand unified 
theory (GUT) scale. At the EW scale the ratio of the $U(1)$ and $SU(2)$ 
gaugino masses are $M_1/M_2 = 1:2$. If in addition, $M_1<\mu$ and the value of 
$\kappa$ is large (so that the effective gauge singlet neutrino mass  
$2 \kappa \nu^c$ is large), the lightest neutralino is essentially bino 
dominated and it is the LSP. First we consider the case when the composition 
of the lightest neutralino is such that, the bino-component $|N_{71}|^2>$ 0.92 
and neutrino masses follow the normal hierarchical pattern. We have observed 
that for the bino dominated case, the lightest neutralino 
(${\tilde \chi}^0_7$) couplings to $\ell^{\pm}$--$W^{\mp}$ pair 
(where $\ell = e, \mu$ or $\tau$) depend on the 
quantities $b_i$ along with a factor which is independent of various lepton 
generations. Naturally, we would expect that the ratios of various decay 
branching ratios such as BR(${\tilde \chi}^0_7 \rightarrow e + W$), 
BR(${\tilde \chi}^0_7 \rightarrow \mu + W$), and  
BR(${\tilde \chi}^0_7 \rightarrow \tau + W$) show nice correlations with the
quantities $b^2_i/b^2_j$ with i,j being $e, \mu$ or $\tau$. This feature is
evident from fig.\ref{nor-bino-bmbebt}. Here we have scanned the parameter 
space of the three neutrino Yukawa couplings with random values for a 
particular choice of the couplings $\lambda$, $\kappa$ and the associated soft 
SUSY breaking trilinear parameters, as well as other MSSM parameters. The 
trilinear soft parameters $A_\nu$ corresponding to $Y_\nu$s also vary  
randomly in a certain range. In addition we have imposed the condition that 
the lightest neutralino (which is the LSP) is bino dominated and neutrino mass 
pattern is normal hierarchical.  

\vspace{0.5cm}
\FIGURE{\epsfig{file=Figures/nnBDBRmt.eps,height=5.00cm} 
\vspace{0.65cm}
\epsfig{file=Figures/nnBDBRet.eps,height=5.00cm} 
\vspace{0.65cm}
\epsfig{file=Figures/nnBDBRem.eps,height=5.00cm} 
\caption{Ratio $\frac{Br(\chi^0_7 \longrightarrow \ell_i~W)}{Br(\chi^0_7
\longrightarrow \ell_j~W)}$ versus $\frac{b^2_i}{b^2_j}$ plot for a bino like
lightest neutralino (the LSP) with bino component, $|N_{71}|^2>$~0.92, where 
$i,j,k~=~e,\mu,\tau$.  Neutrino mass pattern is taken to be normal 
hierarchical. Choice of parameters are $M_1=110$ GeV, $\lambda=0.13, 
\kappa=0.65, m_{\tilde{\nu}^c}=300~{\rm GeV} ~{\rm and} ~m_{\tilde{L}}=400 
~{\rm GeV}$. Mass of the LSP is 106.9 GeV. 
The value of the $\mu$ parameter comes out to be -228.9 GeV.}
\label{nor-bino-bmbebt}}
\FIGURE{\epsfig{file=Figures/Tsq23BDnn.eps,height=5.00cm} 
\vspace{0.5cm}
\epsfig{file=Figures/Tsq13BDnn.eps,height=5.00cm} 
\caption{Ratio $\frac{Br(\chi^0_7 \longrightarrow \mu~W)}{Br(\chi^0_7
\longrightarrow \tau~W)}$ versus $\tan^2\theta_{23}$~(left),
$\frac{Br(\chi^0_7 \longrightarrow e~W)}{\sqrt{Br(\chi^0_7 \longrightarrow
\mu~W)^2+Br(\chi^0_7 \longrightarrow \tau~W)^2}}$ with $\tan^2\theta_{13}$~(right) plot for a bino dominated lightest neutralino (the LSP) with bino component, $|N_{71}|^2>$~0.92. Neutrino mass pattern is normal hierarchical. Choice of parameters are same as that of fig.7.}
\label{nor-bino-LSP}}

We have checked that the correlations between the ratios of the lightest 
neutralino decay branching ratios and $b^2_i/b^2_j$ is more prominent 
with increasing bino component of the lightest neutralino. Note that when 
$(b_i/b_j)^2\rightarrow 1$ the ratios of branching ratios shown in 
fig.\ref{nor-bino-bmbebt} also tend to $1$. We have seen earlier that the 
neutrino mixing angles $\theta_{23}$ and $\theta_{13}$ also show nice 
correlation with the ratios $b^2_\mu/b^2_\tau$ and $b^2_e/b^2_\tau$, 
respectively (see Figs.\ref{sinsqtheta23_bmubtau} and 
\ref{sinsqtheta13_bsqratios}). Hence we would expect that the ratios of the 
branching ratios $\frac{{\rm BR}({\tilde \chi}^0_7 \rightarrow \mu W)}
{{\rm BR}({\tilde \chi}^0_7 \rightarrow \tau W)}$ and 
$\frac{BR({\tilde \chi}^0_7 \longrightarrow e~W)}
{\sqrt{BR({\tilde \chi}^0_7 \longrightarrow \mu~W)^2
+BR({\tilde \chi}^0_7 \longrightarrow \tau~W)^2}}$ show correlations with 
$\tan^2\theta_{23}$ and $\tan^2\theta_{13}$. These correlations are shown in
fig.\ref{nor-bino-LSP}. We have seen earlier (see eq.(\ref{eigenvalues})) that
with the normal hierarchical pattern of the neutrino masses, the atmospheric
mass scale is determined by the quantity $\Omega_b = \sqrt{b^2_e + b^2_\mu 
+ b^2_\tau}$. Naturally one would expect that the atmospheric and the reactor
angles are correlated with the $\ell + W$ final states of the lightest 
neutralino decays and no correlation is expected for the solar angle. This is
what we have observed numerically. Here we have considered the regions of the
parameter space where the neutrino mass-squared differences and mixing angles
are within the 3$\sigma$ allowed range shown in 
eqs.(\ref{eq-msq21})--(\ref{eq-mixing-angles}). fig.\ref{nor-bino-LSP} also 
shows the model prediction for the ratios of branching ratios where the 
neutrino experimental data are satisfied. For our sample choice of parameters
in fig.\ref{nor-bino-LSP}, one would expect that the ratio 
$\frac{BR({\tilde \chi}^0_7 \longrightarrow \mu~W)}{BR({\tilde \chi}^0_7
\longrightarrow \tau~W)}$ should be in the range 0.45 to 1.25. Similarly, 
the other ratio $\frac{BR({\tilde \chi}^0_7 \longrightarrow e~W)}
{\sqrt{BR({\tilde \chi}^0_7 \longrightarrow \mu~W)^2
+BR({\tilde \chi}^0_7 \longrightarrow \tau~W)^2}}$ is expected in this case 
to be less than 0.07. We can also see from fig.\ref{nor-bino-LSP} that the 
ratio of branching ratios in the ($\mu + W$) and ($\tau + W$) channels becomes 
almost equal for the maximal value of the atmospheric mixing angle 
($\theta_{23}~=~45^\circ$). On the other hand, we do not observe any 
correlation with the solar mixing angle $\theta_{12}$ since it is a 
complicated function of $a^2_i$ and $b^2_i$ (see eq. (\ref{solar_analytical})).

\FIGURE{\epsfig{file=Figures/niBDBRmt.eps,height=5.00cm} 
\vspace{0.65cm}
\epsfig{file=Figures/niBDBRet.eps,height=5.00cm} 
\vspace{0.65cm}
\epsfig{file=Figures/niBDBRem.eps,height=5.00cm} 
\caption{Ratio $\frac{BR({\tilde \chi}^0_7 \longrightarrow \ell^-_i~W)}
{BR({\tilde \chi}^0_7 \longrightarrow \ell^-_j~W)}$ versus 
$\frac{b^2_i}{b^2_j}$ plot for a bino like lightest neutralino (the LSP) 
with bino component $|N_{71}|^2>$~0.95, where $i,j,k~=~e,\mu,\tau$. Neutrino 
mass pattern is inverted hierarchical. Choice of parameters are 
$M_1=105$ GeV, $\lambda=0.15, \kappa=0.65, m_{\tilde{\nu}^c}=300~{\rm GeV} 
~{\rm and} ~m_{\tilde{L}}=445~{\rm GeV}$. Mass of the LSP is 103.3 GeV.
The value of the $\mu$ parameter comes out to be -263.7 GeV.}
\label{inv-bino-bmbebt}}

In the case of inverted hierarchical mass pattern of the light neutrinos, the
${\tilde \chi}^0_7$--$\ell_i$--$W$ coupling is still controlled by the 
quantities $b^2_i$. Hence the ratios of the branching ratios discussed earlier,
show nice correlations with $b^2_i/b^2_j$ (see fig.\ref{inv-bino-bmbebt}). 
However, in this case the solar mixing angle shows 
some correlation with the ratio $\frac{BR({\tilde \chi}^0_7 
\longrightarrow e~W)} {\sqrt{BR({\tilde \chi}^0_7 \longrightarrow \mu~W)^2
+BR({\tilde \chi}^0_7 \longrightarrow \tau~W)^2}}$. This is shown in 
fig.\ref{inv-bino-LSP}. The correlation is not very sharp and some dispersion 
occurs due to the fact that the two heavier neutrino masses controlling the 
atmospheric mass scale and solar mass-squared difference are not completely 
determined by the quantities $b^2_i$ and there is some contribution of the 
quantities $a^2_i$, particularly for the second heavy neutrino mass eigenstate.
\FIGURE{\epsfig{file=Figures/Tsq12BDni.eps,height=5.00cm} 
\vspace{0.5cm}
\epsfig{file=Figures/Tsq23BDni.eps,height=5.00cm} 
\caption{Ratio $\frac{BR({\tilde \chi}^0_7 \longrightarrow e~W)}
{\sqrt{BR({\tilde \chi}^0_7 \longrightarrow \mu~W)^2+BR({\tilde \chi}^0_7 
\longrightarrow \tau~W)^2}}$ with $\tan^2\theta_{12}$ (left) plot for a bino 
dominated lightest neutralino (LSP) with bino component $|N_{71}|^2>$~0.95. 
In the right figure the ratio $\frac{BR({\tilde \chi}^0_7 \longrightarrow 
\mu~W)}{BR({\tilde \chi}^0_7 \longrightarrow \tau~W)}$ versus 
$\tan^2\theta_{23}$ is plotted.  Neutrino mass pattern is assumed to be 
inverted hierarchical. Choice of parameters are same as that of 
fig.9.}
\label{inv-bino-LSP}}
The correlation of the ratio $\frac{BR({\tilde \chi}^0_7 \longrightarrow 
\mu~W)} {BR({\tilde \chi}^0_7 \longrightarrow \tau~W)}$ with 
$\tan^2\theta_{23}$ shows a different behaviour compared to what we have seen
in the case of normal hierarchical scenario. This is because in the case of 
inverted hierarchical mass pattern of the neutrinos, $\tan^2\theta_{23}$ 
decreases with increasing $b^2_\mu/b^2_\tau$. One can observe from 
Figs.\ref{nor-bino-LSP} and \ref{inv-bino-LSP} that if the experimental 
value of the ratio $\frac{BR({\tilde \chi}^0_7 \longrightarrow e~W)}
{\sqrt{BR({\tilde \chi}^0_7 \longrightarrow \mu~W)^2+BR({\tilde \chi}^0_7
\longrightarrow \tau~W)^2}}$ is $\ll$ 1 then that indicates a normal 
hierarchical neutrino mass pattern for a bino-dominated lightest neutralino 
LSP whereas a higher value ($\sim$ 1) of this ratio measured in experiments
might indicate that the neutrino mass pattern is inverted hierarchical. 
Similarly a measurement of the ratio $\frac{BR({\tilde \chi}^0_7 
\longrightarrow \mu~W)}{BR({\tilde \chi}^0_7 \longrightarrow \tau~W)}$ can
also give an indication regarding the particular hierarchy of the neutrino
mass pattern in the case of a bino dominated LSP. 

\subsubsection{Higgsino dominated lightest neutralino}
When one considers higher values of the $U(1)$ gaugino mass $M_1$, i.e. 
$M_1>\mu$ and large value of $\kappa$ (so that the effective gauge singlet 
neutrino mass $2 \kappa \nu^c$ is large), the lightest neutralino is  
essentially higgsino dominated and it is the LSP. Naturally one needs to 
consider a small value of the coupling $\lambda$ so that the effective $\mu$ 
parameter ($\mu = 3 \lambda \nu^c$) is smaller. In order to look at the 
lightest neutralino decay branching ratios in this case, we consider a 
situation where the higgsino component in ${\tilde \chi}^0_7$ is 
$|N_{73}|^2 + |N_{74}|^2 >$ 0.90. As in the case of a bino dominated LSP,
the generation dependence of the ${\tilde \chi}^0_7$--$\ell_i$--$W$ couplings
comes through the quantities $b^2_i$. However, because of the large value of 
the $\tau$ Yukawa coupling, the higgsino--$\tau$ mixing is larger and as a 
result the partial decay width of ${\tilde \chi}^0_7$ into ($W + \tau$) is 
larger than into ($W + \mu$) and ($W + e$). This feature is shown in 
fig.\ref{nor-higgsino-bmbebt}, where the ratios of branching ratios are 
plotted against the quantities $b^2_i/b^2_j$. The domination of
BR(${\tilde \chi}^0_7 \rightarrow \tau + W$) over the other two is clearly
evident. Nevertheless, all the three ratios of branching ratios show 
sharp correlations with the corresponding $b^2_i/b^2_j$. In this figure
the normal hierarchical pattern of the neutrino masses has been considered.
As in the case of a bino LSP, here also the ratios $\frac{BR({\tilde \chi}^0_7 
\longrightarrow \mu~W)} {BR({\tilde \chi}^0_7 \longrightarrow \tau~W)}$ and
$\frac{BR({\tilde \chi}^0_7 \longrightarrow e~W)}
{\sqrt{BR({\tilde \chi}^0_7 \longrightarrow \mu~W)^2+BR({\tilde \chi}^0_7
\longrightarrow \tau~W)^2}}$ show nice correlations with neutrino mixing
angles $\theta_{23}$ and $\theta_{13}$, respectively. This is shown in 
fig.\ref{nor-higgsino-LSP}. However, in this case the predictions for these
two ratios are very different from the bino LSP case. The expected value of
the ratio $\frac{BR({\tilde \chi}^0_7 \longrightarrow \mu~W)} 
{BR({\tilde \chi}^0_7 \longrightarrow \tau~W)}$ is approximately between 
0.05 and 0.10 in a region where one can accommodate the experimental 
neutrino data. Similarly, the predicted value of the ratio 
$\frac{BR({\tilde \chi}^0_7 \longrightarrow e~W)}
{\sqrt{BR({\tilde \chi}^0_7 \longrightarrow \mu~W)^2+BR({\tilde \chi}^0_7
\longrightarrow \tau~W)^2}}$ is $\leq$ 0.006. On the other hand, there is 
no such correlations with the solar mixing angle $\theta_{12}$.

\vspace{0.5cm}
\FIGURE{\epsfig{file=Figures/nnHDBRmt.eps,height=5.00cm} 
\vspace{0.65cm}
\epsfig{file=Figures/nnHDBRet.eps,height=5.00cm} 
\vspace{0.65cm}
\epsfig{file=Figures/nnHDBRem.eps,height=5.00cm} 
\caption{Ratio $\frac{BR({\tilde \chi}^0_7 \longrightarrow l_i~W)}
{BR({\tilde \chi}^0_7 \longrightarrow l_j~W)}$ versus $\frac{b^2_i}{b^2_j}$ 
plot for a higgsino like LSP with higgsino component $(|N_{73}|^2+
|N_{74}|^2)>$~0.95, where $i,j,k~=~e,\mu,\tau$. Neutrino mass pattern is 
normal hierarchical. Choice of parameters are $M_1=325$ GeV, $\lambda=0.06, 
\kappa=0.65, m_{\tilde{\nu}^c}=300~{\rm GeV} ~\rm{and} ~m_{\tilde{L}}=
400~{\rm GeV}$. Mass of the LSP is 98.6 GeV.
The value of the $\mu$ parameter comes out to be -105.9 GeV.}
\label{nor-higgsino-bmbebt}}
\vspace{0.5cm}
\FIGURE{\epsfig{file=Figures/Tsq23HDnn.eps,height=5.00cm} 
\vspace{0.5cm}
\epsfig{file=Figures/Tsq13HDnn.eps,height=5.00cm} 
\caption{Ratio $\frac{BR({\tilde \chi}^0_7 \longrightarrow \mu~W)}
{BR({\tilde \chi}^0_7 \longrightarrow \tau~W)}$ versus $\tan^2\theta_{23}$~
(left), $\frac{BR({\tilde \chi}^0_7 \longrightarrow e~W)}
{\sqrt{BR({\tilde \chi}^0_7 \longrightarrow \mu~W)^2+
BR({\tilde \chi}^0_7 \longrightarrow \tau~W)^2}}$ with $\tan^2\theta_{13}$
~(right) plot for a higgsino LSP with higgsino component  
$(|N_{73}|^2+|N_{74}|^2)>$~0.95. Neutrino mass pattern is normal 
hierarchical. Choice of parameters are same as that of 
fig.11.}
\label{nor-higgsino-LSP}}
\vspace{2.3cm}

Similar correlations of the ratios of branching ratios with $b^2_i/b^2_j$ 
are also obtained for a higgsino dominated LSP in the case where the neutrino 
mass pattern is inverted hierarchical. Once again it shows that the 
${\tilde \chi}^0_7$ decays to ($\tau + W$) channel is dominant over the 
channels ($e + W$) and ($\mu + W$) for any values of $b^2_i/b^2_j$ because 
of the larger $\tau$ Yukawa coupling. On the other hand, the correlations with
the neutrino mixing angles show a behaviour similar to that of a bino LSP
with inverted neutrino mass hierarchy though with much smaller values for the
ratios $\frac{BR({\tilde \chi}^0_7 \rightarrow \mu~W)}
{BR({\tilde \chi}^0_7 \rightarrow \tau~W)}$ 
and $\frac{BR({\tilde \chi}^0_7 \rightarrow e~W)}
{\sqrt{BR({\tilde \chi}^0_7 \rightarrow \mu~W)^2+Br({\tilde \chi}^0_7 
\rightarrow \tau~W)^2}}$. These are shown in fig.\ref{inv-higgsino-LSP}.
Note that the correlations in this case are not very sharp, especially with 
$\tan^2\theta_{12}$. Thus we see that small values of these ratios 
(for both normal and inverted hierarchy) are characteristic features of a 
higgsino dominated LSP in this model.

\subsubsection{$\nu^c$ dominated lightest neutralino}
Because of our choice of parameters i.e., a generation independent coupling 
$\kappa$ of the gauge singlet neutrinos and a common VEV $\nu^c$,
the three neutralino mass eigenstates which are predominantly gauge singlet 
neutrinos are essentially mass degenerate. There is a very small mass 
splitting due to mixing. However, unlike the case of a bino or higgsino 
dominated lightest neutralino, these $\nu^c$ dominated lightest neutralino 
states cannot be considered as the LSP. This is because in this case the 
lightest scalar (which is predominantly a gauge singlet sneutrino 
${\tilde \nu}^c$) is the lightest supersymmetric particle. This is very 
interesting since usually one does not get a ${\tilde \nu}^c$ as an LSP 
in a model where the gauge singlet neutrino superfield has a large Majorana 
mass term in the superpotential. However, in this case the effective Majorana 
mass term is at the EW scale and there is also a contribution from the 
trilinear scalar coupling $A_\kappa \kappa$ which keeps the mass of the 
singlet scalar sneutrino smaller. It is also very interesting to study the 
decay patterns of the lightest neutralino in this case since here one can 
probe the gauge singlet neutrino mass scales at the colliders. 
\vspace{1.7cm}
\FIGURE{\epsfig{file=Figures/Tsq23HDni.eps,height=4.50cm} 
\vspace{0.5cm}
\epsfig{file=Figures/Tsq12HDni.eps,height=4.60cm} 
\caption{Ratio $\frac{BR({\tilde \chi}^0_7 \longrightarrow \mu~W)}
{BR({\tilde \chi}^0_7 \longrightarrow \tau~W)}$ versus $\tan^2\theta_{23}$
~(left), $\frac{BR({\tilde \chi}^0_7 \longrightarrow e~W)}
{\sqrt{BR({\tilde \chi}^0_7 \longrightarrow \mu~W)^2+Br({\tilde \chi}^0_7 
\longrightarrow \tau~W)^2}}$ with $\tan^2\theta_{12}$~(right) plot
for a higgsino LSP with higgsino component $(|N_{73}|^2+|N_{74}|^2)>$~0.95. 
Neutrino mass pattern is inverted hierarchical. Choice of parameters are 
$M_1=490$ GeV, $\lambda=0.07, \kappa=0.65, m_{\tilde{\nu}^c}=320 \rm{GeV} 
~\rm{and} ~m_{\tilde{L}}=430 \rm{GeV}$. Mass of the LSP is 110.8 GeV.
The value of the $\mu$ parameter comes out to be -115.3 GeV.}
\label{inv-higgsino-LSP}}

Before discussing the decay patterns of the lightest neutralino which is 
$\nu^c$ dominated, let us say a few words regarding their production at 
the LHC. The direct production of $\nu^c$ (by $\nu^c$ we mean the $\nu^c$
dominated lightest neutralino in this section) is negligible because of
the very small mixing with the MSSM neutralinos. Nevertheless, they can be 
produced at the end of the cascade decay chains of the squarks and gluinos at
the LHC. For example, if the next-to-next-to-lightest SUSY particle (NNLSP) 
is higgsino dominated (this is the state above the three almost degenerate 
lightest neutralinos) and it has a non-negligible mixing with $\nu^c$ (remember 
that the higgsino--$\nu^c$ mixing occurs mainly because of the 
term $\lambda {\hat \nu}^c {\hat H}_1 {\hat H}_2$ in the superpotential), then 
the branching ratio of the decay ${\tilde H} \rightarrow Z + \nu^c$ can be 
larger than the branching ratios in the $\ell W$ and $\nu Z$ channels. This way one can produce $\nu^c$ dominated lightest neutralino. Similarly, a higgsino dominated lighter chargino can also produce gauge singlet neutrinos. Another way of producing 
$\nu^c$ is through the decay of an NNLSP ${\tilde \tau}_1$, such as 
${\tilde \tau}_1 \rightarrow \tau + \nu^c$. A detailed discussion of these 
issues is beyond the scope of the present paper and we hope to come back to 
this in a future publication \cite{pradipta-sr-collider}. 
      
\vspace{0.5cm}
\FIGURE{\epsfig{file=Figures/nnNCDBRmt_old.eps,height=5.00cm} 
\vspace{0.5cm}
\epsfig{file=Figures/nnNCDBRmt.eps,height=5.00cm} 
\caption{Ratio $\frac{BR({\tilde \chi}^0_7 \longrightarrow \mu~W)}
{BR({\tilde \chi}^0_7 \longrightarrow \tau~W)}$ versus 
$\frac{a^2_\mu}{a^2_\tau}$ (left) and versus $\frac{b^2_\mu}{b^2_\tau}$ 
(right) plot for a $\nu^c$ like lightest neutralino (${\tilde \chi}^0_7$) 
with $\nu^c$ component $(|N_{75}|^2+|N_{76}|^2+|N_{77}|^2)>$~0.99, (left) and
$>$0.97 (right). Neutrino mass pattern is normal hierarchical. Choice of 
parameters are for (left) $M_1=405$ GeV, $\lambda=0.29, \kappa=0.07, 
(A_\lambda \lambda)= -8.2~{\rm TeV} \times \lambda,(A_\kappa \kappa)= 165 
~{\rm GeV} \times \kappa,~m_{\tilde{\nu}^c}=50~{\rm GeV} ~\rm{and} 
~m_{\tilde{L}}=825~{\rm GeV}$ and for (right) $M_1=405$ GeV, $\lambda=0.10, 
\kappa=0.07, (A_\lambda \lambda)= -2~{\rm TeV} \times \lambda,(A_\kappa \kappa)
= 165~{\rm GeV} \times \kappa, ~m_{\tilde{\nu}^c}=50~{\rm GeV} ~{\rm and} 
~m_{\tilde{L}}=825~{\rm GeV}$. Mass of the lightest neutralino is 129.4 GeV 
(left) and 119.8 GeV (right) respectively.
The values of the $\mu$ parameter are -803.9 GeV and -258.8 GeV, respectively.}
\label{nor-singlino-bmbt-amat}}

When one considers higher value of the gaugino mass, i.e. $M_1>\mu$ and 
a small value of the coupling $\kappa$ (so that the effective Majorana mass of 
$\nu^c$ is small, i.e. $2 \kappa \nu^c < \mu$), the lightest neutralino is 
essentially $\nu^c$ dominated. As we have mentioned earlier, in this case the 
LSP is the scalar partner of $\nu^c$, i.e. ${\tilde \nu}^c$. However, the 
decay of $\nu^c$ into $\nu + {\tilde \nu}^c$ is suppressed compared to the 
decays $\nu^c \rightarrow \ell_i + W$ and $\nu^c \rightarrow \nu_i +Z$ that 
we have considered so far. Because of this in this section we will neglect the
decay $\nu^c \rightarrow \nu + {\tilde \nu}^c$ while discussing the correlation
of the lightest neutralino (${\tilde \chi}^0_7$) decays with the neutrino 
mixing angles.

In this case the coupling of the lightest neutralino (${\tilde \chi}^0_7$) 
with $\ell_i$--$W$ pair depends on the $\nu^c$ content of ${\tilde \chi}^0_7$. 
Note that the $\nu^c$ has a very small mixing with the MSSM neutralino states. 
However, in some cases the $\nu^c$ dominated lightest neutralino can have a 
non-negligible higgsino component. In such cases the coupling 
${\tilde \chi}^0_7$--$\ell_i$--$W$ depends mainly on the quantities $b_i$. On 
the other hand, if ${\tilde \chi}^0_7$ is very highly dominated by $\nu^c$, 
then the coupling ${\tilde \chi}^0_7$-- $\ell_i$--$W$ has a nice correlation 
with the quantities $a_i$. So in order to study the decay correlations of the 
$\nu^c$ dominated lightest neutralino, we consider two cases (i) $\nu^c$ 
component is $>$ 0.99, and (ii) $\nu^c$ component is $>0.97$ with some 
non-negligible higgsino admixture.

The correlations of the decay branching ratio 
$\frac{BR({\tilde \chi}^0_7 \longrightarrow \mu~W)}
{BR({\tilde \chi}^0_7 \longrightarrow \tau~W)}$ are shown in 
fig.\ref{nor-singlino-bmbt-amat} for the cases (i) and (ii) mentioned above. 
As we have explained already, this figure demonstrates that in case (i) the
ratio of the branching ratio is dependent on the quantity $a^2_\mu/a^2_\tau$
whereas in case (ii) this ratio is correlated with $b^2_\mu/b^2_\tau$ though
there is some suppression due to large $\tau$ Yukawa coupling. 
\vspace{0.5cm}
\FIGURE{\epsfig{file=Figures/nnNCDBRem_old.eps,height=5.00cm} 
\vspace{0.5cm}
\epsfig{file=Figures/nnNCDBRem.eps,height=5.00cm} 
\caption{Ratio $\frac{BR({\tilde \chi}^0_7 \longrightarrow e~W)}
{BR({\tilde \chi}^0_7 \longrightarrow \mu~W)}$ versus 
$\frac{a^2_e}{a^2_\mu}$ (left) and versus $\frac{b^2_e}{b^2_\mu}$ 
(right) plot for a $\nu^c$ like lightest neutralino (${\tilde \chi}^0_7$) 
with $\nu^c$ component $(|N_{75}|^2+|N_{76}|^2+|N_{77}|^2)>$~0.99 (left), and
$>$0.97 (right). Neutrino mass pattern is normal hierarchical. Choice of 
parameters are same as that of fig.14.}
\label{nor-singlino-bebm-aeam}}

Similar calculations were performed also for the other ratios discussed
earlier. For example, in fig.\ref{nor-singlino-bebm-aeam} we have shown the 
variations of the ratio $\frac{BR({\tilde \chi}^0_7 \longrightarrow e~W)} 
{BR({\tilde \chi}^0_7 \longrightarrow \mu~W)}$ as functions of 
$\frac{a^2_e}{a^2_\mu}$ and $\frac{b^2_e}{b^2_\mu}$ for the cases (i) and (ii),
respectively. The variation with $\frac{a^2_e}{a^2_\mu}$ is not sharp and 
dispersive in nature whereas the variation with $\frac{b^2_e}{b^2_\mu}$ is 
very sharp and shows that in this case the relevant couplings are proportional 
to $b_e$ and $b_\mu$, respectively. 
\vspace{0.5cm}
\FIGURE{\epsfig{file=Figures/Tsq23NCDnn_old.eps,height=5.00cm} 
\vspace{0.5cm}
\epsfig{file=Figures/Tsq12NCDnn.eps,height=5.00cm} 
\caption{Ratio $\frac{BR({\tilde \chi}^0_7 \longrightarrow \mu~W)}
{BR({\tilde \chi}^0_7 \longrightarrow \tau~W)}$ versus 
$\tan^2\theta_{23}$~(left), $\frac{BR({\tilde \chi}^0_7 \longrightarrow e~W)}
{\sqrt{BR({\tilde \chi}^0_7 \longrightarrow \mu~W)^2
+BR({\tilde \chi}^0_7 \longrightarrow \tau~W)^2}}$ with $\tan^2\theta_{12}$
~(right) plot for a $\nu^c$ dominated lightest neutralino with $\nu^c$ component 
$(|N_{75}|^2+|N_{76}|^2+|N_{77}|^2)>$~0.99 (left) and $>$ 0.97 (right). 
Neutrino mass pattern is normal hierarchical. Choice of parameters are same as that of fig.14.}
\label{nor-singlino-LN}}

On the other hand, in case (i) only $\tan^2\theta_{23}$ shows a nice 
correlation with the ratio $\frac{BR({\tilde \chi}^0_7 \longrightarrow \mu~W)}
{BR({\tilde \chi}^0_7 \longrightarrow \tau~W)}$ (see 
fig.\ref{nor-singlino-LN}) and $\tan^2\theta_{12}$ or $\tan^2\theta_{13}$ 
does not show any correlation with the other ratio. The non-linear behaviour 
of the ratios of branching ratios in case(i) is due to the fact that the
parameters $Y_\nu$s (which control the $a_i$) appear both in the neutralino
and chargino mass matrices. The charged lepton Yukawa couplings also play a 
role in determining the ratios. One can also see that the
prediction for this ratio of branching ratio for case (i), as shown in 
fig.\ref{nor-singlino-LN}, is in the range 0.5--3.5, which is larger compared 
to the bino dominated or higgsino dominated cases for both normal and 
inverted hierarchical pattern of neutrino masses. Also, the nature of this 
variation is similar to what we see with the inverted hierarchical pattern of 
neutrino masses in the bino or higgsino dominated cases. 

In case (ii) none of the neutrino mixing angles show very good correlations 
with the ratios of branching ratios that we have been discussing. However, one 
can still observe some kind of a correlation of the ratio 
$\frac{BR({\tilde \chi}^0_7 \longrightarrow e~W)} {\sqrt{BR({\tilde \chi}^0_7 
\longrightarrow \mu~W)^2 +BR({\tilde \chi}^0_7 \longrightarrow \tau~W)^2}}$ 
with $\tan^2\theta_{12}$. The prediction for this ratio from the neutrino data
is on the smaller side ($\sim$ 0.07). 
\vspace{0.5cm}
\FIGURE{\epsfig{file=Figures/niNCDBRmt_old.eps,height=5.00cm} 
\vspace{0.5cm}
\epsfig{file=Figures/niNCDBRmt.eps,height=5.00cm} 
\caption{Ratio $\frac{BR({\tilde \chi}^0_7 \longrightarrow \mu~W)}
{BR({\tilde \chi}^0_7 \longrightarrow \tau~W)}$ versus 
$\frac{a^2_\mu}{a^2_\tau}$ (left) and versus $\frac{b^2_\mu}{b^2_\tau}$ 
(right) plot for a $\nu^c$ like lightest neutralino (${\tilde \chi}^0_7$) 
with $\nu^c$ component $(|N_{75}|^2+|N_{76}|^2+|N_{77}|^2)>$~0.99 (left), and
$>$0.97 (right). Neutrino mass pattern is inverted hierarchical. Choice of 
parameters are for (left) $M_1=445$ GeV, $\lambda=0.29, \kappa=0.07, 
(A_\lambda \lambda)= -8.2~{\rm TeV} \times \lambda,(A_\kappa \kappa)= 
165~{\rm GeV} \times \kappa,~m_{\tilde{\nu}^c}=50~{\rm GeV} ~{\rm and} 
~m_{\tilde{L}}=835 \rm{GeV}$ and for (right) $M_1=445$ GeV, $\lambda=0.10, 
\kappa=0.07, (A_\lambda \lambda)= -2~{\rm TeV} \times \lambda,(A_\kappa \kappa)
= 165~{\rm GeV} \times \kappa, ~m_{\tilde{\nu}^c}=50~{\rm GeV} ~{\rm and} 
~m_{\tilde{L}}=835~{\rm GeV}$. Mass of the lightest neutralino is 129.4 GeV 
(left) and 119.8 GeV (right) respectively.}
\label{inv-singlino-bmbt-amat}}
\vspace{2.3cm}

With the inverted hierarchical neutrino mass pattern, in case (i)
one observes a sharp correlation of the ratio 
$\frac{BR({\tilde \chi}^0_7 \longrightarrow \mu~W)}
{BR({\tilde \chi}^0_7 \longrightarrow \tau~W)}$ with 
$\frac{a^2_\mu}{a^2_\tau}$ (see fig.\ref{inv-singlino-bmbt-amat}). The other 
two ratios $\frac{BR({\tilde \chi}^0_7 \longrightarrow e~W)}
{BR({\tilde \chi}^0_7 \longrightarrow \mu~W)}$ and 
$\frac{BR({\tilde \chi}^0_7 \longrightarrow e~W)}
{BR({\tilde \chi}^0_7 \longrightarrow \tau~W)}$ do not show very sharp 
correlations with $\frac{a^2_e}{a^2_\mu}$ and $\frac{a^2_e}{a^2_\tau}$,
respectively and we do not plot them here. However, in case (ii) all the
three ratios show nice correlations with the corresponding $b^2_i/b^2_j$. We
have shown this in fig.\ref{inv-singlino-bmbt-amat} only for 
$b^2_\mu/b^2_\tau$. In this case the variations of the ratios of branching 
ratios with neutrino mixing angles are shown in fig.\ref{inv-singlino-LN}. 

\vspace{0.9cm}
\FIGURE{\epsfig{file=Figures/Tsq23NCDni.eps,height=5.00cm} 
\vspace{0.5cm}
\epsfig{file=Figures/Tsq12NCDni.eps,height=5.00cm} 
\caption{Ratio $\frac{BR({\tilde \chi}^0_7 \longrightarrow \mu~W)}
{BR({\tilde \chi}^0_7 \longrightarrow \tau~W)}$ versus $\tan^2\theta_{23}$
~(left), $\frac{BR({\tilde \chi}^0_7 \longrightarrow e~W)}
{\sqrt{BR({\tilde \chi}^0_7 \longrightarrow \mu~W)^2
+BR({\tilde \chi}^0_7 \longrightarrow \tau~W)^2}}$ with $\tan^2\theta_{12}$
~(right) plot for a $\nu^c$ dominated lightest neutralino with $\nu^c$ 
component $(|N_{75}|^2+|N_{76}|^2+|N_{77}|^2)>$~0.97. Neutrino mass pattern is 
inverted hierarchical. Choice of parameters are same as that of fig.17.}
\label{inv-singlino-LN}}
For the case (i), only $\tan^2\theta_{13}$ shows certain correlation with the
ratio of branching ratio shown in fig.\ref{inv-singlino-LN} (right), but we
do not show it here.

Finally, we would like to mention that in all these different cases discussed 
above, the lightest neutralino can have a finite decay length which can produce
displaced vertices in the vertex detectors. Depending on the composition of the
lightest neutralino, one can have different decay lengths. For example, a
bino-dominated lightest neutralino can produce a displaced vertex $\sim$ a few
mm. Similarly, for a higgsino dominated lightest neutralino, decay vertices of
the order of a few cms can be observed. On the other hand, if the lightest 
neutralino is $\nu^c$ dominated, then the decay lengths can be of the order 
of a few meters. These are very unique predictions of this model which can, 
in principle, be tested at the LHC.    


\section{Summary and Conclusion}
\label{Summary and conclusion}
In this work we have studied a supersymmetric model in detail where the 
observed pattern of neutrino mass-squared differences and mixing angles are 
obtained with the help of three standard model gauge-singlet neutrino 
superfields, which simultaneously solve the $\mu$ problem of MSSM. The 
additional terms in the superpotential and the scalar potential include 
R-parity violating interactions involving these gauge-singlet neutrino 
superfields. The vacuum expectation values of the singlet sneutrinos give 
rise to effective Majorana mass terms for the singlet neutrinos, as well as a 
$\mu$-term, both at the electroweak scale. This model was introduced in 
ref.\cite{munoz-lopez_fogliani} and some phenomenology was discussed for a 
single gauge-singlet neutrino superfield. The spectrum and parameter space of 
this model, with three gauge singlet neutrino superfields, were discussed in
\cite{munoz-lopez-2}. We have performed a detailed and extensive analysis of 
this model in the neutrino and neutralino sector with the inclusion of three 
generation of gauge singlet neutrino superfields along with the associated 
interaction terms. The neutrino mass matrix is obtained because of the 
electroweak scale seesaw mechanism involving the gauge-singlet neutrinos and 
the mixing between the MSSM neutralinos and the neutrinos. We have done 
a thorough and systematic study of the neutrino mass matrix both analytically 
and numerically and tried to identify 
the relevant parameters which crucially control the bilarge pattern of 
neutrino mixing angles. We show that even with a flavour diagonal structure of 
the neutrino Yukawa coupling matrix, two large and one small mixing angles can 
be generated in this case. Both the normal and inverted hierarchical pattern 
of neutrino masses can be obtained with different hierarchies of the neutrino 
Yukawa couplings. Because of the presence of the neutrino-neutralino mixing, 
it is in general difficult to obtain a degenerate mass spectrum of the 
neutrinos and we do not consider this possibility in this paper. 

We have also looked at the scalar sector of this model and wrote down the
neutral scalar, pseudoscalar and charged scalar mass-squared matrices 
of this model. The allowed regions in the model parameters are obtained which 
satisfy certain constraints in the scalar sector. For example, absence of 
tachyons in the scalar squared-mass eigenvalues puts severe constraints in the 
relevant parameter space. With these choices of parameters, satisfying the 
scalar sector constraints, we have tried to fit the global three flavour 
neutrino data with both normal and inverted hierarchical mass spectrum of the 
neutrinos. Since this model involves several free parameters, we have made a 
few simplifying assumptions which makes this model more predictive.

Perhaps the most interesting part of this whole analysis is the study of the
phenomenology of the lightest neutralino which can be the LSP in some cases or 
it can be the NLSP. The decay patterns of the lightest neutralino may provide 
additional information to find out more about the neutrino mass patterns and 
mixing angles. We have considered three different scenarios where the lightest
neutralino can be either a bino-dominated one or a higgsino-dominated one or 
it can be mostly a gauge-singlet neutrino with very little mixing with the 
other states. We have also discussed briefly the production mechanism of the
gauge-singlet neutrino dominated lightest neutralino at the LHC. The study of
the decay pattern and the production mechanism of the lightest neutralino in 
this last mentioned scenario is extremely important because it will help in 
probing the mass scale and the properties of the gauge-singlet neutrinos at the
LHC. The presence of the gauge-singlet neutrino dominated lightest neutralino
can also distinguish this model from the usual bilinear R-parity violating 
model of generating neutrino masses and mixing. An important test of this 
model, as a supersymmetric solution to the observed neutrino mass patterns 
and mixing, can be performed by measuring the ratios of the decay branching 
ratios of the lightest neutralino in the final states involving a charged 
lepton and a W-boson. We have shown explicitly that these ratios of branching 
ratios have certain correlations with the neutrino mixing angles which depend 
on the nature of the LSP as well as on the pattern of the neutrino mass 
hierarchies considered. The study of the higgsino dominated LSP case is also 
very important because it can provide information about the $\mu$ parameter 
which is determined in terms of the vacuum expectation values of the 
gauge-singlet sneutrinos. Thus one may have information about the 
gauge-singlet neutrino mass scale and its coupling, from the decay pattern of 
the higgsino dominated LSP. Collider phenomenology of this model at the LHC is 
very rich both in the fermionic and the scalar sector. For example, pair 
produced lightest neutralino at the LHC give rise to the final states 
$\mu\mu WW$, $\tau\tau WW$ and $\mu\tau WW$ with a certain ratio for their 
production rates nicely correlated with $\tan^2\theta_{23}$. These production 
rates also depend on the dominant component of the lightest neutralino as well
as on the different hierarchical patterns of the three light neutrino masses. 
Another important testable prediction of this model is the measurement of 
displaced vertices in the decay of the lightest neutralino. This decay length 
can vary in the range of a few mm to $\sim$ 1 meter depending on the nature of 
the lightest neutralino. We hope to come back to these issues in a future 
publication \cite{pradipta-sr-collider}.

\vspace{1cm}

\noindent {\bf Acknowledgments} 

\noindent PG would like to thank the Council of Scientific and Industrial 
Research, Govt. of India for the financial support received as a Junior 
Research Fellow. We thank U. Chattopadhyay, D. Das, B. Mukhopadhyaya and 
S.K. Rai for many useful comments and suggestions. We would also like to 
thank K. Ray for a very useful suggestion during the developement of the code 
for numerical calculations. 



\appendix

 \section{Scalar mass squared matrices}\label{appenxA}
In this appendix we present the details of various scalar mass squared 
matrices. For the convenience of the reader let us repeat that the scalar 
mass squared matrices are now $8\times8$, considering all three generations of 
sneutrinos (both doublet and singlet) and charged sleptons. These enhancements 
are essentially due to the mixing of neutral Higgs bosons with both the 
doublet and singlet sneutrinos and the mixing of charged Higgs with the 
charged sleptons.

\subsection{Neutral scalar mass squared matrices}
\label{Neutral scalar mass squared matrices}

The decomposition of various neutral scalar fields in 
real(${\mathcal R}$) and imaginary(${\mathcal I}$) parts are as follows
\bea
{H^0_1}&=&{H^0_{1{\mathcal R}}}+i{H^0_{1{\mathcal I}}},\nonumber \\
{H^0_2}&=&{H^0_{2{\mathcal R}}}+i{H^0_{2{\mathcal I}}}, \nonumber \\
{\tilde \nu}^c_k&=&{\tilde \nu}^c_{k{\mathcal R}} 
+i{\tilde \nu}^c_{k{\mathcal I}},\nonumber \\
{\tilde \nu}_k&=&{\tilde \nu}_{k{\mathcal R}}+i{\tilde \nu}_{k{\mathcal I}}.
\nonumber \\
\label{decomposition}
\eea
Only the real components get VEVs as indicated in eq.(\ref{vevs}).  

The entries of the scalar and pseudoscalar mass-squared matrices are defined as
\bea
({M^2_P})^{\alpha \beta}&=&\langle\frac{1}{2}\frac{\partial^2{V_{neutral}}}
{\partial{\phi^{\alpha}_{\mathcal I}}\partial{\phi^{\beta}_{\mathcal I}}}
\rangle, \nonumber \\
({M^2_S})^{\alpha \beta}&=&\langle\frac{1}{2}\frac{\partial^2{V_{neutral}}}
{\partial{\phi^{\alpha}_{\mathcal R}}\partial{\phi^{\beta}_{\mathcal R}}}
\rangle,
\label{mass_square_matrix_working_formula}
\eea
where 
\bea
{\phi^{\alpha}_{\mathcal I}}&=&{H^0_{1{\mathcal I}}}, {H^0_{2{\mathcal I}}}, 
{\tilde \nu}^c_{k{\mathcal I}}, {\tilde \nu}_{k{\mathcal I}}, \nonumber \\
{\phi^{\alpha}_{\mathcal R}}&=&{H^0_{1{\mathcal R}}}, {H^0_{2{\mathcal R}}}, 
{\tilde \nu}^c_{k{\mathcal R}}, {\tilde \nu}_{k{\mathcal R}}.
\eea
Note that the Greek indices $\alpha, \beta$ are used to refer 
various scalar and pseudoscalar Higgs and both doublet and singlet 
sneutrinos, that is $H^0_1, H^0_2, {\tilde \nu}^c_k, {\tilde \nu}_k$, 
whereas k is used as a subscript to specify various flavours of doublet and 
singlet sneutrinos i.e., $k=e, {\mu}, {\tau}$ in the flavour 
(weak interaction) basis.

\subsubsection{CP-odd neutral mass squared matrix}
\label{CP-odd neutral mass squared matrix}

The basis for CP-odd or pseudoscalar mass-squared matrix is
\bea\label{pseudoscalar_basis}
\Phi^T_{P}=(H^0_{1{\mathcal I}},H^0_{2{\mathcal I}},
{\tilde \nu}^c_{n{\mathcal I}},{\tilde \nu}_{n{\mathcal I}}).
\eea
The pseudoscalar mass term in the Lagrangian is of the form
\beq\label{pseudoscalar_Lagrangian}
{\mathcal{L}_{pseudoscalar}^{mass}} = {\Phi_{P}^T} \mathcal{M}^2_{P} {\Phi_{P}},
\eeq
where $\mathcal{M}^2_{P}$ is an $8\times8$ symmetric matrix. Using 
eq.(\ref{Minim3}), (\ref{Minim4}), and eq.(\ref{Abbrevations}), the 
independent elements are given by

\bea
(M^2_P)^{H^0_{1{\mathcal I}} H^0_{1{\mathcal I}}}&=&{\frac{1}{v_1}}
\left[{\sum_j}{\lambda^j}{v_2} \left({\sum_{ik}} {\kappa^{ijk}}{\nu^c_i}
{\nu^c_k}\right)+{\sum_j} {\lambda^j}{r^j}{v^2_2}+ {\mu}{\sum_j}{r^j_c}{\nu_j}
+{\sum_i} (A_{\lambda}{\lambda})^{i}{\nu^c_i}{v_2} \right], \nonumber \\
(M^2_P)^{H^0_{1{\mathcal I}} H^0_{2{\mathcal I}}}&=&{\sum_{i,j,k}}{\lambda^j}
{\kappa^{ijk}} {\nu^c_i}{\nu^c_k} + \sum_i (A_{\lambda} {\lambda})^{i}{\nu^c_i},
\nonumber \\
(M^2_P)^{H^0_{2{\mathcal I}} H^0_{2{\mathcal I}}}&=&{\frac{1}{v_2}}
\left[-{\sum_{j}}{\rho^{j}} \left({\sum_{l,k}} \kappa^{ljk}{\nu^c_l}
{\nu^c_k}\right)-{\sum_{i,j}} (A_{\nu}{Y_{\nu}})^{ij} {\nu_i}{\nu^c_j}
+{\sum_i}(A_\lambda {\lambda})^{i} {\nu^c_i}{v_1}\right], 
\nonumber \\
(M^2_P)^{H^0_{1{\mathcal I}} {\tilde \nu}^c_{m{\mathcal I}}}&=&-2 {\sum_j}
{\lambda^j}{u^{mj}_c}{v_2} - {\mu}{r^m} + {\lambda^m}{\sum_i}{r^i_c}{\nu_i} 
+ (A_{\lambda}{\lambda})^m {v_2}, \nonumber \\
(M^2_P)^{H^0_{1{\mathcal I}} {\tilde \nu}_{m{\mathcal I}}}&=&-{\sum_j}
{\lambda^j}{Y^{mj}_{\nu}} {v^2_2} - {\mu}{r^m_c},\nonumber \\
(M^2_P)^{H^0_{2{\mathcal I}} {\tilde \nu}^c_{m{\mathcal I}}}&=&2{\sum_j}
{u^{mj}_c}{\rho^j} - {\sum_i}(A_{\nu} Y_{\nu})^{im}{\nu_i} + 
(A_{\lambda}{\lambda})^m {v_1}, \nonumber \\
(M^2_P)^{H^0_{2{\mathcal I}} {\tilde \nu}_{m{\mathcal I}}}&=&-{\sum_{i,j,k}}
{Y^{mj}_{\nu}} {\kappa^{ijk}} {\nu^c_i}{\nu^c_k}-{\sum_j}
({A_{\nu}}{Y_{\nu}})^{mj}{\nu^c_j}, 
\nonumber \\
(M^2_P)^{{\tilde \nu}^c_{n{\mathcal I}} {\tilde \nu}^c_{m{\mathcal I}}}&=&
-2{\sum_j}{\kappa^{jnm}} {\zeta^j}+4{\sum_j} {u^{mj}_c}{u^{nj}_c}+{\rho^m}
{\rho^n}+{h^{nm}}{v^2_2} + (m^2_{\tilde{\nu}^c})^{nm} - 2{\sum_{i}} 
({A_{\kappa}}{\kappa})^{inm} {\nu^c_i},\nonumber \\
(M^2_P)^{{\tilde \nu}^c_{n{\mathcal I}} {\tilde \nu}_{m{\mathcal I}}}&=&
2{\sum_j}{u^{mj}_c} {Y^{nj}_{\nu}}{v_2}-{Y^{nm}_{\nu}}{\sum_i}{r^i_c}{\nu_i}
+{r^n_c}{r^m} +{\mu}{v_1}{Y^{nm}_{\nu}} -{\lambda^m}{r^n_c}{v_1}
-{({A_{\nu}}{Y_{\nu}})^{nm}} {v_2},\nonumber \\
(M^2_P)^{{\tilde \nu}_{n{\mathcal I}} {\tilde \nu}_{m{\mathcal I}}}&=&
{\sum_j}{Y^{mj}_{\nu}} {Y^{nj}_{\nu}}{v^2_2} +{r^m_c}{r^n_c}
+(m^2_{\tilde{L}})^{nm}+{\gamma_g} {\xi_{\upsilon}} {\delta_{mn}},\nonumber \\
\label{element_of_pseudoscalar_mass_matrix}
\eea

where
\beq\label{pseudoscalar_mass_sq_matrix_assumption}
h^{nm}={\lambda^n}{\lambda^m}+{\sum_{i}}{Y_{\nu}^{in}} {Y_{\nu}^{im}}.
\eeq
We have checked that one eigenvalue of this $8\times8$ matrix is zero 
corresponding to the neutral Goldstone boson.

\subsubsection{CP-even neutral mass squared matrix}
\label{CP-even neutral mass squared matrix}
The basis for the CP-even or scalar mass-squared matrix is

\beq\label{scalar_basis}
\Phi^T_{S}=({H^0_{1{\mathcal R}}},{H^0_{2{\mathcal R}}},
{{\tilde{\nu}}^c_{n{\mathcal R}}},{{\tilde{\nu}_{n{\mathcal R}}}}).
\eeq

The scalar mass term in the Lagrangian is of the form

\beq\label{scalar_Lagrangian}
{\mathcal{L}_{scalar}^{mass}} = {\Phi_{S}^T} \mathcal{M}^2_{S} {\Phi_{S}},
\eeq
where $\mathcal{M}^2_{S}$ is an $8\times8$ symmetric matrix. The independent 
entries using eq.(\ref{Minim3}), (\ref{Minim4}), and eq.(\ref{Abbrevations}) 
are given by

\bea
(M^2_S)^{H^0_{1{\mathcal R}} H^0_{1{\mathcal R}}}&=&{\frac{1}{v_1}}\left[{\sum_j}{\lambda^j}{v_2}
\left({\sum_{ik}} {\kappa^{ijk}}{\nu^c_i}{\nu^c_k}\right)+{\sum_j}{\lambda^j}
{r^j}{v^2_2}+{\mu} {\sum_j}{r^j_c}{\nu_j}+{\sum_i}(A_{\lambda}{\lambda})^{i}
{\nu^c_i}{v_2}\right] +2{\gamma_g}{v^2_1}, \nonumber \\
(M^2_S)^{H^0_{1{\mathcal R}} H^0_{2{\mathcal R}}}&=&-2{\sum_j}{\lambda^j}{\rho^j}{v_2}-{\sum_{i,j,k}}{\lambda^j}
{\kappa^{ijk}}{\nu_i}{\nu^c_k}-2{\gamma_g}{v_1}{v_2}-{\sum_i}{({A_{\lambda}}
{\lambda})^i}{\nu^c_i}, \nonumber\\
(M^2_S)^{H^0_{2{\mathcal R}} H^0_{2{\mathcal R}}}&=&{\frac{1}{v_2}}\left[-{\sum_{j}}{\rho^{j}}\left({\sum_{l,k}}
\kappa^{ljk}{\nu^c_l}{\nu^c_k}\right)-{\sum_{i,j}}(A_{\nu}{Y_{\nu}})^{ij}
{\nu_i}{\nu^c_j}+{\sum_i}(A_\lambda {\lambda})^{i} {\nu^c_i}{v_1}\right]
+2{\gamma_g}{v^2_2},\nonumber \\
(M^2_S)^{H^0_{1{\mathcal R}} {\tilde{\nu}^c_{m{\mathcal R}}}}&=&-2{\sum_j}{\lambda^j}{u^{mj}_c}{v_2}+2{\mu}
{v_1}{\lambda^m}-{\lambda^m}{\sum_i}{r^i_c}{\nu_i}-{\mu}{r^m}-
{({A_{\lambda}}{\lambda})^m}{v_2}, \nonumber\\
(M^2_S)^{H^0_{1{\mathcal R}} {\tilde{\nu}_{m{\mathcal R}}}}&=&-{\sum_j}{\lambda^j}{Y^{mj}_{\nu}}{v^2_2}-
{\mu}{r^m_c}+2{\gamma_g}{\nu_m}{v_1},\nonumber \\
(M^2_S)^{H^0_{2{\mathcal R}} {\tilde{\nu}^c_{m{\mathcal R}}}}&=&2{\sum_j}{u^{mj}_c}{\rho^j}+2{\lambda^m}
{\mu}{v_2}+2{\sum_i}{Y^{im}_{\nu}}{r^i_c}{v_2}
+{\sum_i}{(A_{\nu}{Y_{\nu}})^{im}}{\nu_i}-{(A_{\lambda}{\lambda})^{m}}{v_1},
\nonumber \\
(M^2_S)^{H^0_{2{\mathcal R}} {\tilde{\nu}_{m{\mathcal R}}}}&=&2{\sum_j}{Y^{mj}_{\nu}}{\rho^j}{v_2}
+{\sum_{i,j,k}}{Y^{mj}_{\nu}}{\kappa^{ijk}}{\nu^c_i}{\nu^c_k}
-2{\gamma_g}{\nu_m}{v_2}+{\sum_j}{(A_{\nu}{Y_{\nu}})^{mj}}{\nu^c_j},
\nonumber \\
(M^2_S)^{{\tilde{\nu}^c_{n{\mathcal R}}} {\tilde{\nu}^c_{m{\mathcal R}}}}&=&2{\sum_j}{\kappa^{jnm}}{\zeta^j}
+4{\sum_j}{u^{mj}_c}{u^{nj}_c}+{\rho^m}{\rho^n}+{h^{nm}}{v^2_2}
+{(m^2_{\tilde{\nu}^c})^{mn}}+2{\sum_i}{(A_{\kappa}{\kappa})^{imn}}{\nu^c_i},
\nonumber \\
(M^2_S)^{{\tilde{\nu}^c_{n{\mathcal R}}} {\tilde{\nu}_{m{\mathcal R}}}}&=&2{\sum_j}{Y^{nj}_{\nu}}{u^{mj}_c}{v_2}
+{Y^{nm}_{\nu}}{\sum_i}{r^i_c}{\nu_i}+{r^n_c}{r^m}-{\mu}{v_1}{Y^{nm}_{\nu}}
-{\lambda^m}{r^n_c}{v_1}+{(A_{\nu}{Y_{\nu}})^{nm}}{v_2},\nonumber \\
(M^2_S)^{{\tilde{\nu}_{n{\mathcal R}}} {\tilde{\nu}_{m{\mathcal R}}}}&=&{\sum_j}{Y^{nj}_{\nu}}{Y^{mj}_{\nu}}{v^2_2}
+{r^m_c}{r^n_c}+{\gamma_g}{\xi_{\upsilon}}{\delta_{nm}}+2{\gamma_g}{\nu_n}
{\nu_m}+{(m^2_{\tilde{L}})^{mn}},\nonumber \\
\label{element_of_scalar_mass_matrix}
\eea
where eq.(\ref{pseudoscalar_mass_sq_matrix_assumption}) has been used.

\subsection{Charged scalar mass squared matrix}
\label{Charged scalar mass squared matrix}

The charged scalar mass squared matrix considering all three generations 
of both left handed and right handed charged sleptons is also an 
$8\times8$ matrix. For the sake of completeness, here we give the expressions
for various elements of the charged scalar mass-squared matrix. 
The elements of the charged scalar mass-squared matrix are defined as

\beq
({M^2_C})^{\alpha \beta} = \langle\frac{1}{2}\frac{\partial^2{V_{charged}}}
{\partial{\phi^{c^\alpha}}\partial{\phi^{c^\beta}}}\rangle,
\label{charged_scalar_mass_square_matrix_working_formula}
\eeq
where 
\beq
{\phi^{c^\alpha}}={H^+_{1}}, {H^+_{2}}, {{\tilde{e}^+_{Rk}}}, 
{{\tilde{e}^+_{Lk}}}.
\eeq
The basis for charged scalar mass-squared matrix is 
\beq\label{charged-scalar_basis}
\Phi^{+^T}_{C}=({H^+_{1}}, {H^+_{2}}, 
{{\tilde{e}}^+_{Rn}} ,{{\tilde{e}}^+_{Ln}}),
\eeq
and the charged scalar mass term in the Lagrangian is of the form
\beq\label{charged-scalar_Lagrangian}
{\mathcal{L}_{charged~scalar}^{mass}} = {\Phi_{C}^{-^T}} \mathcal{M}^2_{C} 
{\Phi_{C}^{+}},
\eeq
where $\mathcal{M}^2_{C}$ is an $8\times8$ symmetric matrix. The independent 
elements of $\mathcal{M}^2_{C}$ using eqs.(\ref{Minim3}), (\ref{Minim4}), and 
eq.(\ref{Abbrevations}) are given by

\bea
(M^2_C)^{H_{1} H_{1}}&=&{\frac{1}{v_1}}\left[{\sum_j} \lambda^j \zeta^j v_2 
+ \mu {\sum_j} r^j_c \nu_j + {\sum_i} ( A_{\lambda} \lambda )^i \nu^c_i v_2 
\right] +  {\sum_{i,j,k}} Y^{ij}_e Y^{kj}_e \nu_i \nu_k - \frac{{g_2}^2}{2} 
({\sum_i} \nu^2_i - v^2_2 ), \nonumber \\
(M^2_C)^{H_{1} H_{2}}&=& -{\sum_j} \lambda^{j^2} v_1 v_2 + {\sum_j} \lambda^j r^j v_2 
+ {\sum_j} \lambda^j u^{ij}_c \nu^c_i + \frac{{g_2}^2}{2} v_1 v_2 + {\sum_i} 
(A_{\lambda} \lambda )^i \nu^c_i, \nonumber\\
(M^2_C)^{H_{2} H_{2}}&=&{\frac{1}{v_2}}\left[ - {\sum_j} \rho^j  \zeta^j  
- {\sum_{i,j}} ( A_\nu Y_\nu )^{ij} \nu_i \nu^c_j + {\sum_i} ( A_\lambda 
\lambda )^i \nu^c_i v_1\right] + \frac{{g_2}^2}{2} ( {\sum_i} \nu^2_i + v^2_1 ),
\nonumber \\
(M^2_C)^{H_{1} {{\tilde{e}}_{Rm}}}&=& - {\sum_i} r^i_c Y^{im}_e v_2 - {\sum_i} 
( A_e Y_e )^{im} \nu_i, \nonumber\\
(M^2_C)^{H_{1} {{\tilde{e}}_{Lm}}}&=& - \mu r^m_c - {\sum_{i,j}} Y^{mj}_e Y^{ij}_e 
\nu_i v_1 + \frac{g_2^2}{2} \nu_m v_1,\nonumber \\
(M^2_C)^{H_{2} {{\tilde{e}}_{Rm}}}&=& - \mu {\sum_i} Y^{mi}_e \nu_i - {\sum_i} 
Y^{im}_e r^i_c v_1,\nonumber \\
(M^2_C)^{H_{2} {{\tilde{e}}_{Lm}}}&=& - {\sum_j} Y^{mj}_{\nu} \zeta^j  + 
\frac{g_2^2}{2} \nu_m v_2 - {\sum_i} (A_{\nu} Y_{\nu})^{mi} \nu^c_i,\nonumber \\
(M^2_C)^{{{\tilde{e}}_{Rn}} {{\tilde{e}}_{Rm}}}&=& {\sum_{i,j}} Y^{im}_e Y^{jn}_e \nu_i \nu_j 
+ {\sum_i} Y^{im}_e Y^{in}_e v^2_1 + (m^2_{\tilde{e}^c})^{mn} - 
\frac{g_1^2}{2} {\xi_\upsilon} \delta_{mn},\nonumber \\
(M^2_C)^{{{\tilde{e}}_{Rn}} {{\tilde{e}}_{Lm}}}&=& - \mu Y^{mn}_e v_2 + (A_e Y_e)^{nm} v_1 ,
\nonumber \\
(M^2_C)^{{{\tilde{e}}_{Ln}} {{\tilde{e}}_{Lm}}}&=& r^m_c r^n_c + {\sum_j} Y^{mj}_e Y^{nj}_e 
v^2_1 + \gamma_g {\xi_\upsilon} \delta_{mn} -\frac{g_2^2}{2}{\xi_\upsilon} 
\delta_{mn} + \frac{g_2^2}{2} \nu_m \nu_n + (m^2_{\tilde{L}})^{mn}.
\label{element_of_charged-scalar_mass_matrix}
\eea

As mentioned earlier, we have computed the eigenvalues of the charged scalar 
mass-squared matrix numerically and ensured that seven of its eigenvalues are 
positive and there is a charged Goldstone boson. 

 \section{Feynman rules}\label{appenxB}
In this appendix we will study the relevant Feynman rules for the LSP decay 
calculations \cite{Haber-Kane, Gunion-Haber-1}. The required Feynman rules are
\begin{enumerate}
\item Neutralino-neutralino - $Z^0$,
\item Chargino-neutralino - $W^{\pm}$.
\end{enumerate}



\subsection{Neutralino-neutralino-$Z^0$ and chargino-chargino-$Z^0$,$\gamma$}

For neutralinos the following relations between mass and weak eigenstates are 
very useful

\bea\label{neutralino_mass-basis_weak-basis_relations}
& &{P_L}{\tilde{B}}^0 = P_L N^*_{i1} \tilde{\chi}^0_i, \nonumber \\
& &{P_R}{\tilde{B}}^0 = P_R N_{i1} \tilde{\chi}^0_i, \nonumber \\
& &{P_L}{\tilde{W}}^0_3 = P_L N^*_{i2} \tilde{\chi}^0_i, \nonumber \\
& &{P_R}{\tilde{W}}^0_3 = P_R N_{i2} \tilde{\chi}^0_i, \nonumber \\
& &{P_L}{\tilde{H}}_j = P_L N^*_{i,j+2} \tilde{\chi}^0_i, \nonumber \\
& &{P_R}{\tilde{H}}_j = P_R N_{i,j+2} \tilde{\chi}^0_i, \quad \text{where} \quad 
j = 1,2,\nonumber \\
& &{P_L}{\nu}^k_L = P_L N^*_{i,k+7} \tilde{\chi}^0_i, \nonumber \\
& &{P_R}{\nu}^k_L = P_R N_{i,k+7} \tilde{\chi}^0_i, \nonumber \\
& &{P_L}{\nu}^{c^k}_R = P_L N^*_{i,k+4} \tilde{\chi}^0_i, \nonumber \\
& &{P_R}{\nu}^{c^k}_R = P_R N_{i,k+4} \tilde{\chi}^0_i, \quad \text{where} \quad 
k = 1,2,3,
\eea
with i varies from 1 to 10 and 
\beq
P_{L}=\left(\frac{1-{\gamma^5}}{2}\right), \quad 
P_{R}=\left(\frac{1+{\gamma^5}}{2}\right).
\eeq

In terms of the four component spinors $\chi_i$ for charginos, the following 
relations between mass and weak eigenstates are very useful

\bea\label{chargino_mass-basis_weak-basis_relations}
& &{P_L}{\tilde{W}} = P_L V^*_{i1} \tilde{\chi}_i, \nonumber \\
& &{P_R}{\tilde{W}} = P_R U_{i1} \tilde{\chi}_i, \nonumber \\
& &{P_L}{\tilde{H}} = P_L V^*_{i2} \tilde{\chi}_i, \nonumber \\
& &{P_R}{\tilde{H}} = P_R U_{i2} \tilde{\chi}_i,\nonumber \\
& &{P_L}{l^j} = P_L V^*_{i,j+2} \tilde{\chi}_i, \nonumber \\
& &{P_R}{l^j} = P_R U_{i,j+2} \tilde{\chi}_i,
\eea
where $j = 1,2,3$, and i varies from 1 to 5.

So in terms of physical or mass eigenstates of charginos and neutralinos the 
required interaction terms are as follows

\bea\label{mass eigenstates of chargino and neutralino}
\mathcal{L}_{Z^0 \tilde{\chi}\tilde{\chi}} &=& \left(\frac{g}{\cos\theta_W}
\right) Z_{\mu} {\bar{\tilde\chi}^+_i} \gamma^{\mu} \left[ O^{\prime^L}_{ij} 
P_L + O^{\prime^R}_{ij} P_R \right] {{\tilde\chi}^+_j} \nonumber \\
& & + \left(\frac{g}{2 \cos\theta_W}\right) Z_{\mu} {\bar{\tilde\chi}^0_i} 
\gamma^{\mu} \left[ O^{{\prime\prime}^L}_{ij} P_L + O^{{\prime\prime}^R}_{ij} 
P_R \right] {{\tilde\chi}^0_j}, \nonumber \\
\mathcal{L}_{\gamma {\tilde{\chi}^+} {\tilde{\chi}^-}} &=& - e A_{\mu} 
{\bar{\tilde\chi}^+_i} \gamma^{\mu} {\tilde\chi}^+_i,
\eea
where
\bea\label{specifications_of_symbols_used_in_mass_eigenstates_of_CN_and_NN}
O^{\prime^L}_{ij} &=& - V_{i1} V^*_{j1} - \frac{1}{2} V_{i2} V^*_{j2} 
+ \delta_{ij} \sin^2 \theta_W, \nonumber \\
O^{\prime^R}_{ij} &=& - U^*_{i1} U_{j1} - \frac{1}{2} U^*_{i2} U_{j2}  
- \frac{1}{2} U^*_{i,k+2} U_{j,k+2}, + \delta_{ij} \sin^2 \theta_W, \nonumber \\
O^{{\prime\prime}^L}_{ij} &=& - \frac{1}{2} N_{i3} N^*_{j3} + \frac{1}{2} 
N_{i4} N^*_{j4} - \frac{1}{2} N_{i,k+7} N^*_{j,k+7}, \nonumber \\
O^{{\prime\prime}^R}_{ij} &=& - {O^{{\prime\prime}^L}_{ij}}^*, k = 1,2,3. 
\eea

In deriving 
eq.(\ref{specifications_of_symbols_used_in_mass_eigenstates_of_CN_and_NN}) 
unitary properties of U and V matrices has been used.
\FIGURE{\epsfig{file=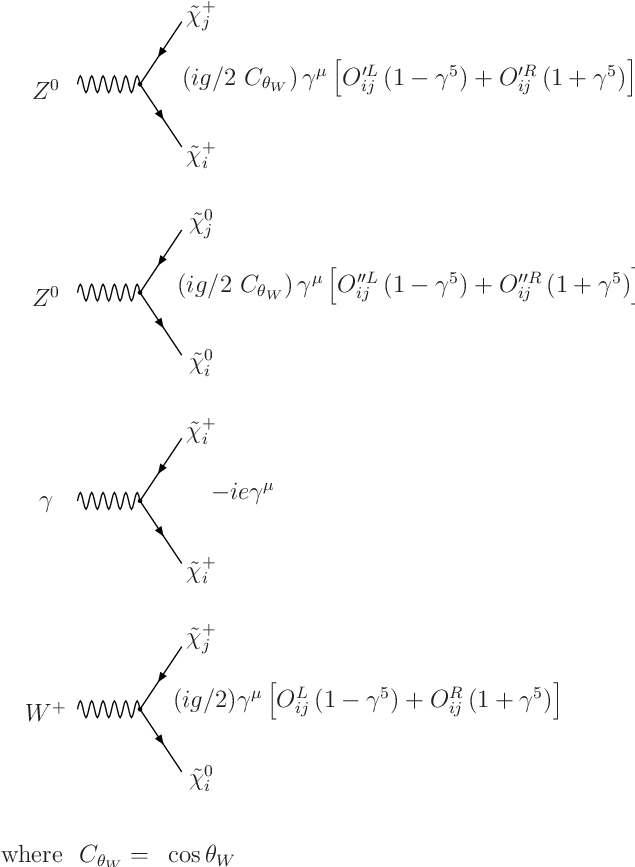,height=12.00cm} 
\caption{Feynman rules for interaction of neutral and charged gauge boson with 
charginos and neutralinos.}
\label{Feynman_Diagram}}

\subsection{Chargino-neutralino - $W^{\pm}$}

Now in terms of physical chargino and neutralino states the relevant 
interaction term is 

\beq\label{Lagrangian of chargino neutralino charged gauge Bosons interaction}
\mathcal{L}_{{W^\mp}{\tilde{\chi}^{\pm}}{\tilde{\chi}^0}} = \mathcal{L}_{{W^-}
{\tilde{\chi}^+}{\tilde{\chi}^0}} + \mathcal{L}_{{W^+}{\tilde{\chi}^-}
{\tilde{\chi}^0}},
\eeq
where $\mathcal{L}_{{W^+}{\tilde{\chi}^-}{\tilde{\chi}^0}}$ is the hermitian 
conjugate of $\mathcal{L}_{{W^-}{\tilde{\chi}^+}{\tilde{\chi}^0}}$ and
\bea\label{Lagrangian of chargino neutralino W^- interaction}
\mathcal{L}_{{W^-}{\tilde{\chi}^+}{\tilde{\chi}}^0} = g W^-_{\mu} 
\bar{\tilde{\chi}}^0_i \gamma^{\mu} \left[ O^L_{ij} P_L + O^R_{ij} P_R\right] 
\tilde{\chi}^+_j,
\eea
with
\bea\label{specifications_of_symbols_used_in_mass_eigenstates_of_CN_and_NN2}
& & O^L_{ij} = N_{i2} V^*_{j1} - \frac{1}{\sqrt{2}} N_{i4} V^*_{j2},
\nonumber \\
& & O^R_{ij} = N^*_{i2} U_{j1} + \frac{1}{\sqrt{2}} N^*_{i3} U_{j2} + 
\frac{1}{\sqrt{2}} N^*_{i,k+7} U_{j,k+2},\nonumber \\
\eea
where $k = 1,2,3$.

The Feynman rules are shown in fig.\ref{Feynman_Diagram}. The matrices $O^{\prime L}_{ij}$, 
$O^{\prime R}_{ij}$, $O^{\prime \prime L}_{ij}$, $O^{\prime \prime R}_{ij}$ 
and $O^{L}_{ij}$, $O^{R}_{ij}$ are defined by 
eq.(\ref{specifications_of_symbols_used_in_mass_eigenstates_of_CN_and_NN}) and 
eq.(\ref{specifications_of_symbols_used_in_mass_eigenstates_of_CN_and_NN2}), respectively.


\begin{thebibliography}{999}
\bibitem{boris_review_08}
  See,for example,B.~Kayser,
  {\it{Neutrino Mass, Mixing, and Flavor Change}},
  \arXivid{0804.1497} [hep-ph] and references therein.

\bibitem{Schwetz-Valle}
  T.~Schwetz, M.~Tortola and J.~W.~F.~Valle,
  {\it{Three-flavour neutrino oscillation update}},
  \newjournal{New.\ J.\ Phys.\ }{NJPH}{10}{2008}{113011}
  [\arXivid{0808.2016} [hep-ph]].

\bibitem{r-parity-mssm}
  P.~Fayet,
  {\it{Supergauge Invariant Extension Of The Higgs Mechanism And A Model For The Electron And Its
 Neutrino}},
    \npb{90}{1975}{104};\\
  {\it{Spontaneously Broken Supersymmetric Theories Of Weak, Electromagnetic And Strong
 Interactions}},
    \plb{69}{1977}{489};\\
   G.~R.~Farrar and P.~Fayet,
  {\it{Phenomenology Of The Production, Decay, And Detection Of New Hadronic States Associated With
Supersymmetry}},
    \plb{76}{1978}{575}.


\bibitem{r-parity-mssm-neutrino1}
  C.~S.~Aulakh and R.~N.~Mohapatra,
  {\it{Neutrino As The Supersymmetric Partner Of The Majoron}},
   \plb{119}{1982}{136};\\
    L.~J.~Hall and M.~Suzuki,
  {\it{Explicit R-Parity Breaking In Supersymmetric Models}},
    \npb{231}{1984}{419};\\
    I.~H.~Lee,
  {\it{Lepton Number Violation In Softly Broken Supersymmetry}},
   \plb{138}{1984}{121};\\
   I.~H.~Lee,
  {\it{Lepton Number Violation In Softly Broken Supersymmetry. 2}},
  \npb{246}{1984}{120};\\
  G.~G.~Ross and J.~W.~F.~Valle,
  {\it{Supersymmetric Models Without R-Parity}},
   \plb{151}{1985}{375};\\
     J.~R.~Ellis, G.~Gelmini, C.~Jarlskog, G.~G.~Ross and J.~W.~F.~Valle,
  {\it{Phenomenology Of Supersymmetry With Broken R-Parity}},
   \plb{150}{1985}{142}.

\bibitem{r-parity-mssm-neutrino2}
   See, for example,
      A.~S.~Joshipura and M.~Nowakowski,
  {\it{'Just so' oscillations in supersymmetric standard model}},
   \prd{51}{1995}{2421} [\hepph{9408224}];\\
  {\it{Leptonic CP violation in supersymmetric Standard Model}},
   \prd{51}{1995}{5271} [\hepph{9403349}];\\
     M.~Nowakowski and A.~Pilaftsis,
  {\it{W and Z boson interactions in supersymmetric models with explicit R-parity violation}},
   \npb{461}{1996}{19} [\hepph{9508271}];\\
    F.~Borzumati, Y.~Grossman, E.~Nardi and Y.~Nir,
  {\it{Neutrino masses and mixing in supersymmetric models without R parity}},
   \plb{384}{1996}{123} [\hepph{9606251}];\\
   S.~Roy and B.~Mukhopadhyaya,
  {\it{Some implications of a supersymmetric model with R-parity breaking bilinear interactions}},
   \prd{55}{1997}{7020} [\hepph{9612447}];\\
   K.~Choi, K.~Hwang and E.~J.~Chun,
  {\it{Atmospheric and solar neutrino masses from horizontal U(1) symmetry}},
  \prd{60}{1999}{031301} [\hepph{9811363}];\\
   A.~S.~Joshipura and S.~K.~Vempati,
  {\it{Sneutrino vacuum expectation values and neutrino anomalies through
  trilinear R-parity violation}},
   \prd{60}{1999}{111303} [\hepph{9903435}];\\
    E.~J.~Chun, S.~K.~Kang, C.~W.~Kim and U.~W.~Lee,
  {\it{Supersymmetric neutrino masses and mixing with R-parity violation}},
   \npb{544}{1999}{89} [\hepph{9807327}];\\
     A.~Datta, B.~Mukhopadhyaya and S.~Roy,
  {\it{Constraining an R-parity violating supersymmetric theory from the
  SuperKamiokande data on atmospheric neutrinos}},
  \prd{61}{2000}{055006} [\hepph{9905549}];\\
    M.~Hirsch, M.~A.~Diaz, W.~Porod, J.~C.~Romao and J.~W.~F.~Valle,
  {\it{Neutrino masses and mixings from supersymmetry with bilinear R-parity
  violation: A theory for solar and atmospheric neutrino oscillations}},
   \prd{62}{2000}{113008} [\hepph{0004115}];\\
   \newjournal{Erratum-ibid.\  D}{EBD}{65}{2002}{119901};\\
    M.~A.~Diaz, M.~Hirsch, W.~Porod, J.~C.~Romao and J.~W.~F.~Valle,
  {\it{Solar neutrino masses and mixing from bilinear R-parity broken
  supersymmetry: Analytical versus numerical results}},
   \prd{68}{2003}{013009} [\hepph{0302021}];\\
   \newjournal{Erratum-ibid.\  D}{EBD}{71}{2005}{059904};\\ 
   F.~Takayama and M.~Yamaguchi,
  {\it{Pattern of neutrino oscillations in supersymmetry with bilinear  R-parity violation}},
  \plb{476}{2000}{116} [\hepph{9910320}];\\
   M.~Hirsch, H.~V.~Klapdor-Kleingrothaus and S.~G.~Kovalenko,
  {\it{B-L violating masses in softly broken supersymmetry}},
  \plb{398}{1997}{311} [\hepph{9701253}];\\ 
  Y.~Grossman and H.~E.~Haber,
  {\it{Sneutrino mixing phenomena}},
  \prl{78}{1997}{3438} [\hepph{9702421}];\\ 
  {\it{(S)neutrino properties in R-parity violating supersymmetry. I:
  CP-conserving phenomena}},
  \prd{59}{1999}{093008} [\hepph{9810536}];\\ 
    Y.~Grossman and S.~Rakshit,
  {\it{Neutrino masses in R-parity violating supersymmetric models}},
  \prd{69}{2004}{093002} [\hepph{0311310}]. 
 
\bibitem{r-parity-review}
For reviews on R-parity violation, see, e.g.,
   R.~Barbier {\it et al.},
  {\it{R-parity violating supersymmetry}},
  \prep{420}{2005}{1} [\hepph{0406039}];\\
    M.~Chemtob,
  {\it{Phenomenological constraints on broken R parity symmetry in  supersymmetry models}},
  \ppnp{54}{2005}{71} [\hepph{0406029}].

\bibitem{Kim-Niles} 
  J.~E.~Kim and H.~P.~Nilles,
  {\it{The Mu Problem And The Strong CP Problem}},
  \plb{138}{1984}{150}.

\bibitem{munoz-lopez_fogliani}
   D.~E.~Lopez-Fogliani and C.~Munoz,
  {\it{Proposal for a new minimal supersymmetric standard model}},
   \prl{97}{2006}{041801} [\hepph{0508297}].

  
\bibitem{munoz-lopez-2}
  N.~Escudero, D.~E.~Lopez-Fogliani, C.~Munoz and R.~R.~de Austri,
  {\it{Analysis of the parameter space and spectrum of the $\mu \nu$SSM}},
  \jhep{0812}{2008}{099}  \arXivid{0810.1507} [hep-ph].

  
\bibitem{kitano-oda}
   R.~Kitano and K.~y.~Oda,
  {\it{Neutrino masses in the supersymmetric standard model with right-handed
  neutrinos and spontaneous R-parity violation}},
  \prd{61}{2000}{113001} [\hepph{9911327}].
  
\bibitem{chemtob-pandita}
  P.~N.~Pandita and P.~F.~Paulraj,
  {\it{Infra-red stable fixed points of Yukawa couplings in non-minimal
  supersymmetric standard model with R-parity violation}},
  \plb{462}{1999}{294} [\hepph{9907561}];\\
  P.~N.~Pandita,
  {\it{Nonminimal supersymmetric standard model with baryon and lepton number violation}},
  \prd{64}{2001}{056002} [\hepph{0103005}];\\
   M.~Chemtob and P.~N.~Pandita,
  {\it{Nonminimal supersymmetric standard model with lepton number violation}},
  \prd{73}{2006}{055012} [\hepph{0601159}];\\
  A.~Abada and G.~Moreau,
  {\it{An origin for small neutrino masses in the NMSSM}},
  \jhep{08}{2006}{044} [\hepph{0604216}].
  
\bibitem{gautam-moreau-abada}
  A.~Abada, G.~Bhattacharyya and G.~Moreau,
  {\it{A new mechanism of neutrino mass generation in the NMSSM with broken lepton number}},
  \plb{642}{2006}{503} [\hepph{0606179}].

\bibitem{ellis-gunion-haber}
   J.~R.~Ellis, J.~F.~Gunion, H.~E.~Haber, L.~Roszkowski and F.~Zwirner,
   {\it{Higgs Bosons in a Nonminimal Supersymmetric Model}},
   \prd{39}{1989}{844}.

\bibitem{biswarup-srikanth}
   B.~Mukhopadhyaya and R.~Srikanth,
  {\it{Bilarge neutrino mixing in R-parity violating supersymmetry: The role  of
  right-chiral neutrino superfields}},
  \prd{74}{2006}{075001} [\hepph{0605109}].
  
\bibitem{farzan-valle}
   Y.~Farzan and J.~W.~F.~Valle,
  {\it{R-parity violation assisted thermal leptogenesis in the seesaw mechanism}},
  \prl{96}{2006}{011601} [\hepph{0509280}]. 
  
\bibitem{domain-wall}
  J.~R.~Ellis, K.~Enqvist, D.~V.~Nanopoulos, K.~A.~Olive, M.~Quiros and F.~Zwirner,
  {\it{Problems for (2,0) compactifications}},
  \plb{176}{1986}{403};\\
  B.~Rai and G.~Senjanovic,
  {\it{Gravity and domain wall problem}},
  \prd{49}{1994}{2729} [\hepph{9301240}];\\
  S.~A.~Abel, S.~Sarkar and P.~L.~White,
  {\it{On the Cosmological Domain Wall Problem for the Minimally Extended
  Supersymmetric Standard Model}},
  \npb{454}{1995}{663} [\hepph{9506359}].
  
\bibitem{domain-wall-soln}
  S.~A.~Abel,
  {\it{Destabilising divergences in the NMSSM}},
  \npb{480}{1996}{55} [\hepph{9609323}];\\
  C.~Panagiotakopoulos and K.~Tamvakis,
  {\it{Stabilized NMSSM without domain walls}},
  \plb{446}{1999}{224} [\hepph{9809475}].

\bibitem{math-wolfram}
  Stephen Wolfram, 
  {\it{The Mathematica Book, \tt{5th ed}. (\tt{Wolfram media, 2003})}}.

\bibitem{neutrino_pmns}
  S.~Eidelman {\it et al.} [Particle Data Group],
  {\it{Review of particle physics}},
  \plb{592}{2004}{1}.
  
\bibitem{osc_experiments1}
  S.~Fukuda {\it et al.}  [Super-Kamiokande Collaboration],
  {\it{Tau neutrinos favoured over sterile neutrinos in atmospheric muon  neutrino
  oscillations}},
  \prl{85}{2000}{3999} [\hepex{0009001}];\\
  M.~Ambrosio {\it et al.}  [MACRO Collaboration],
  {\it{Matter effects in upward-going muons and sterile neutrino oscillations}},
  \plb{517}{2001}{59} [\hepex{0106049}];\\
  Q.~R.~Ahmad {\it et al.}  [SNO Collaboration],
  {\it{Direct evidence for neutrino flavour transformation from neutral-current
  interactions in the Sudbury Neutrino Observatory}},
  \prl{89}{2002}{011301} [\nuclex{0204008}];\\
  {\it{Measurement of day and night neutrino energy spectra at SNO and constraints
  on neutrino mixing parameters}},
  \prl{89}{2002}{011302} [\nuclex{0204009}];\\
  S.~N.~Ahmed {\it et al.}  [SNO Collaboration],
  {\it{Measurement of the total active B-8 solar neutrino flux at the Sudbury
  Neutrino Observatory with enhanced neutral current sensitivity}},
  \prl{92}{2004}{181301} [\nuclex{0309004}].

\bibitem{osc_experiments2}
  M.~Apollonio {\it et al.}  [CHOOZ Collaboration],
  {\it{Limits on Neutrino Oscillations from the CHOOZ Experiment}},
  \plb{466}{1999}{415} [\hepex{9907037}];\\
  K.~Eguchi {\it et al.}  [KamLAND Collaboration],
  {\it{First results from KamLAND: Evidence for reactor anti-neutrino disappearance}},
  \prl{90}{2003}{021802} [\hepex{0212021}];\\
  A.~Bandyopadhyay, S.~Choubey, S.~Goswami, S.~T.~Petcov and D.~P.~Roy,
  {\it{Constraints on neutrino oscillation parameters from the SNO salt phase data}},
  \plb{583}{2004}{134} [\hepph{0309174}];\\
  G.~L.~Fogli, E.~Lisi, A.~Marrone, D.~Montanino, A.~Palazzo and A.~M.~Rotunno,
  {\it{Addendum to: Solar neutrino oscillation parameters after first KamLAND results}},
  \prd{69}{2004}{017301} [\hepph{0308055}];\\
  P.~C.~de Holanda and A.~Y.~Smirnov,
  {\it{Solar neutrinos: The SNO salt phase results and physics of conversion}},
  \app{21}{2004}{287} [\hepph{0309299}];\\
  M.~Maltoni, T.~Schwetz, M.~A.~Tortola and J.~W.~F.~Valle,
  {\it{Status of global fits to neutrino oscillations}},
   \newjournal{New.\ J.\ Phys.\ }{NJPH}{6}{2004}{122} [\hepph{0405172}];\\
  A.~Strumia and F.~Vissani,
  {\it{Implications of neutrino data circa 2005}},
  \npb{726}{2005}{294} [\hepph{0503246}].
  
\bibitem{tribimaximal}
  P.~F.~Harrison, D.~H.~Perkins and W.~G.~Scott,
  {\it{Tri-bimaximal mixing and the neutrino oscillation data}},
  \plb{530}{2002}{167} [\hepph{0202074}].

\bibitem{Gunion-Haber-2}
  J.~F.~Gunion and H.~E.~Haber,
  {\it{Two-body decays of neutralinos and charginos}},
  \prd{37}{1988}{2515}.
  
\bibitem{Franke-Frass}
  F.~Franke and H.~Fraas,
  {\it{Neutralinos and Higgs Bosons in the Next-To-Minimal Supersymmetric Standard Model}},
  \ijmpa{12}{1997}{479} [\hepph{9512366}];\\
  {\it{Production and decay of neutralinos in the next-to-minimal  supersymmetric standard model}},
  \newjournal{Z.\ Phys.\  C}{ZPC}{72}{1996}{309} [\hepph{9511275}].
  
\bibitem{correlations_neutralino}
  B.~Mukhopadhyaya, S.~Roy and F.~Vissani,
  {\it{Correlation between neutrino oscillations and collider signals of
  supersymmetry in an R-parity violating model}},
  \plb{443}{1998}{191} [\hepph{9808265}];\\
  E.~J.~Chun and J.~S.~Lee,
  {\it{Implication of Super-Kamiokande data on R-parity violation}},
  \prd{60}{1999}{075006} [\hepph{9811201}];\\
  S.~Y.~Choi, E.~J.~Chun, S.~K.~Kang and J.~S.~Lee,
  {\it{Neutrino oscillations and R-parity violating collider signals}},
  \prd{60}{1999}{075002} [\hepph{9903465}];\\
  J.~C.~Romao, M.~A.~Diaz, M.~Hirsch, W.~Porod and J.~W.~F.~Valle,
  {\it{A supersymmetric solution to the solar and atmospheric neutrino problems}},
  \prd{61}{2000}{071703} [\hepph{9907499}];\\
  W.~Porod, M.~Hirsch, J.~Romao and J.~W.~F.~Valle,
  {\it{Testing neutrino mixing at future collider experiments}},
  \prd{63}{2001}{115004} [\hepph{0011248}].
  
\bibitem{spontaneous-rpv-neutralino}
  M.~Hirsch, A.~Vicente and W.~Porod,
  {\it{Spontaneous R-parity violation: Lightest neutralino decays and neutrino
  mixing angles at future colliders}},
  \prd{77}{2008}{075005} [\arXivid{0802.2896}].

\bibitem{pradipta-sr-collider}
  P. Ghosh and S. Roy, {\it{Signatures of $\mu \nu$SSM at the LHC}}, work in progress.

\bibitem{Haber-Kane}
  H.~E.~Haber and G.~L.~Kane,
  {\it{The Search For Supersymmetry: Probing Physics Beyond The Standard Model}},
  \prep{117}{1985}{75}.

\bibitem{Gunion-Haber-1}  
  J.~F.~Gunion and H.~E.~Haber,
  {\it{Higgs Bosons In Supersymmetric Models. 1}},
  \npb{272}{1986}{1};\\
  \newjournal{Erratum-ibid.\  B}{EBB}{402}{1993}{567};\\
  J.~F.~Gunion and H.~E.~Haber,
  {\it{Higgs Bosons in Supersymmetric Models. 2. Implications for Phenomenology}},
  \npb{278}{1986}{449}.

\end{thebibliography}
\end{document}